\documentclass[proceedings]{JHEP37}
\usepackage{eurosym}
\usepackage{amssymb}
\usepackage{amsfonts,cite}
\usepackage{amsmath}
\usepackage{graphicx}
\usepackage{multirow}
\usepackage{epsfig}
\usepackage{xcolor}

\DeclareMathOperator\sech{sech}

\newbox\mybox

\newcommand\fverb{\setbox\mybox=\hbox\bgroup\verb}
\newcommand\fverbdo{\egroup\medskip\noindent\fbox{\unhbox\mybox}\ }
\newcommand\fverbit{\egroup\item[\fbox{\unhbox\mybox}]}
\conference{Higher time-derivative theories from x-t interchanged integrable field theories}
\abstract{We compare a relativistic and a nonrelativistic version of Ostrogradsky's method for higher-time derivative theories extended to scalar field theories and consider as an alternative a multi-field variant. We apply the schemes to space-time rotated modified Korteweg-de Vries systems and, exploiting their integrability, to Hamiltonian systems built from space-time rotated inverse Legendre transformed higher-order charges of these systems. We derive the equal-time Poisson bracket structures of these theories, establish the integrability of the latter theories by means of the Painlev\'e test and construct exact analytical period benign solutions in terms of Jacobi elliptic functions to the classical equations of motion. The classical energies of these partially complex solutions are real when they respect a certain modified CPT-symmetry and complex when this symmetry is broken. The higher order Cauchy and initial-boundary value problem are addressed analytically and numerically. Finally, we provide the explicit quantization of the simplest mKdV system, exhibiting the usual conundrum of having the choice between either having to deal with a theory that includes non-normalizable states or spectra that are unbounded from below. In our non-Hermitian system the choice is dictated by the correct sign in the decay width.}    

\title{Higher time-derivative theories from space-time interchanged integrable field theories}
\author{Andreas Fring$^\bullet$, Takano Taira$^\circ$ and Bethan Turner$^\bullet$\\
 $\bullet$ Department of Mathematics, City, University of London, Northampton Square,\\ $\,\,$ London EC1V 0HB, UK \\
 $\circ$ Research Fellow of Japan Society for Promotion of Science, Institute of Industrial \\ $\,\,$ Science, The University of Tokyo
 5-1-5 Kashiwanoha,
  Kashiwa 277-8574, Japan\\
 
E-mail: a.fring@city.ac.uk, taira904@iis.u-tokyo.ac.jp, bethan.turner.2@city.ac.uk}

\input{tcilatex}
\begin{document}

\section{Introduction}

Standard quantum field theories only contain derivative terms in their kinetic part up to second order. Ostrogradky's fundamental instability \cite{ostrogradsky1850memoire,Woodard1} is the main reason for this limit, in particular on the higher time-derivatives as their presence will inevitably lead to linear momentum field terms that can not be eliminated by partial integration and subsequent dropping of surface terms. While the presence of these terms is highly unappealing, since at least classically they lead to instabilities that can be reached in finite time, higher time-derivative theories (HTDT) posses also many very attractive features such as for instance being renormalizable {\color{ black} \cite{stelle77ren,grav1,grav2,grav3,modesto16super}.}

These latter properties have nurtured the hope that one might eventually overcome the problematic issues and exploit the positive features. Several proposals of dealing with the deficiencies have been made such as for instance the introduction of constraints \cite{ghostconst}, a Dirac-Pauli quantization scheme with an indefinite metric \cite{salvio16quant},  the introduction of additional degrees of freedom that do not belong to the physical spectrum, so-called fakeons \cite{fakeons}, or the complex extensions of the models to a ${\cal PT}$-symmetric non-Hermitian system \cite{bender2008no,raidal2017quantisation}. 

A useful insight is gained by separating out theories that are already very problematic on the classical level and making a distinction between 
ghost states of malevolent and benign nature. On the classical level the malevolent states are identified as solutions that reach singularities in finite time, whereas benign solutions are oscillatory or might only diverge in infinite time   \cite{smilga2017rev,smilga2021exactly,Smilga6,smilga2021benign}.

Leaving the problematic features largely aside, HTDT appeared in a number of different contexts, such as in attempts to quantise gravity when adding curvature squared terms to the Einstein-Hilbert action \cite{Hawking} or the resolution of the cosmological singularity problem \cite{biswas2010towards}. Finite temperature physics may be formulated in terms of HTDT \cite{weldon98finite} and also
 HTDT black hole solutions have been constructed \cite{mignemi1992black}. Their BRST symmetries have been identified \cite{rivelles2003triviality,Kap1} in some quantum field theories, they were used in a massless particle descriptions of bosons and fermions \cite{plyush89mass,Mpl}, used in the description of the Higgs sector in the standard model and also some supersymmetric versions have been investigated \cite{dine1997comments,smilga17ultrav}. Classical and quantum stability properties of HTDT were investigated in \cite{Sugg1,Sugg2,Sugg3,Sugg4,deffayet22ghost,deffayet23global}. 
 
 The theories studied in detail so far are often simply ad hoc examples or tailored for the specific contexts they are considered in as mentioned above. A systematic study of larger classes of HTDT remains an open issue. Recently Smilga  \cite{smilga2021exactly} pointed out that higher charges of integrable systems naturally involve terms with higher derivatives and when interpreted as Hamiltonians could be suitable candidates to identify classes of theories with benign sectors in their parameter space. It turns out that even in their original form these theories have not been studied systematically, but exhibit interesting features, as seen in higher charge Hamiltonian systems of affine Toda lattice theories type \cite{bethanAF} and scattering theories for multi-particle Calogero and Calogero-Moser systems \cite{fring23int}.
 The HTDT can be obtained from higher charges of these theories by interchanging space and time. Some properties of models obtained in this manner are trivially invariant when $x$ has been interchanged with $t$. However, time is being kept as the flow parameter so that for instance equal-time Poisson bracket change, the Cauchy and the initial-boundary value problem are altered as they require more independent boundary functions etc. As a specific class of integrable systems we investigate here in more detail space-time rotated modified Korteweg-de Vries (mKdV) systems as HTDT.
 
 Our manuscript is organised as follows: In section 2 a comparison between different variants of Ostrogradsky's classical method adapted to scalar field theories is carried out. We consider an alternative equivalent scheme in form of a multi-field theory in which all higher time-derivatives are hidden and set up the procedure on how to build HTDT Hamiltonian systems from higher charges of integrable systems.  In section 3 we apply the schemes to mKdV systems initially for generic values of $n$ for which we derive the equal-time Poisson bracket relations. Subsequently we derive separately for the KdV ($n=3$) and the standard mKdV ($n=4$) the higher charge Hamiltonians, convert the entire theory into a multi-field theory that hides the higher derivatives and establish their integrability by means of the Painlev\'e test. In section 4 we construct exact, partially complex, periodic solutions of benign nature to the equations of motion in terms of Jacobi elliptic functions. We calculate their classical energies, which turn out to be real when they are invariant under certain ${\cal CPT}$-symmetries and complex when this symmetry is broken. We demonstrate that the Cauchy and the initial-value boundary problem is not solved with standard solutions for the $n=2$-theory, and identify some sectors for the $n=3,4$-theories where this is possible. In section 5 we carry out the quantization of the $n=2$-theory. Our conclusions are stated in section 6.

\section{Ostrogradsky's method for scalar field theories}

In this section we compare three versions of Ostrogradsky's method for HTDT extended to scalar field theories: a relativistic, a nonrelativistic, and a multi-field variant. In each of them we recall and elaborate on the procedures to derive the Hamiltonian densities from a given Lagrangian density involving higher derivative terms in the time variable $t$ by employing Ostrogradsky's classical scheme \cite{ostrogradsky1850memoire,Woodard1} adapted to scalar field theories. 

\subsection{Lorentz invariant formulation}

We start with a Lorentz invariant formulation as outlined in \cite{motohashi20q,Urries,thibes21nat} for relativistic field theories that generalise Ostrogradsky's method further from quantum mechanics to field theories. We consider a generic Lagrangian density depending on a scalar field $\varphi$ and its higher derivatives with respect to space and time
\begin{equation}
	{\cal L}(\varphi, \varphi_\mu, \ldots , \varphi_{\mu_1, \ldots , \mu_m}), \qquad \mu_i = x,t
\end{equation}
with $m$ denoting the highest order of the derivatives. Our metric is taken to be flat. The generalised momenta derived from ${\cal L}$ fall into two different types as
\begin{equation}
	\pi^{\mu_1 \dots \mu_m} := \frac{\partial  {\cal L} }{\partial \varphi_{\mu_1 \dots \mu_m}  }, \qquad
	\pi^{\mu_1 \dots \mu_i} := \frac{\partial  {\cal L} }{\partial \varphi_{\mu_1 \dots \mu_i}  }   - \partial_{\mu_{i+1}} 	\pi^{\mu_1 \dots \mu_i \mu_{i+1} },   \quad i=1,\ldots, m-1.  \label{genmom}
\end{equation}
The Legendre transformation in terms of the canonical momenta then defines the Hamiltonian density as
\begin{equation}
	{\cal H}= \pi^\mu \varphi_\mu + \dots + \pi^{\mu_1 \ldots \mu_{m-1} } \varphi_{\mu_1 \ldots \mu_{m-1} }
	+ \pi^{\mu_1 \ldots \mu_{m} } \hat{\varphi}_{\mu_1 \ldots \mu_{m} } - {\cal L}(\varphi, \varphi_\mu, \ldots , \hat{\varphi}_{\mu_1, \ldots , \mu_m}),  \label{Hamildens}
\end{equation}
where $\hat{\varphi}_{\mu_1, \ldots , \mu_m}$ is to be understood as replacing  
\begin{equation}
	\varphi_{\mu_1, \ldots , \mu_m}  \rightarrow \hat{\varphi}_{\mu_1, \ldots , \mu_m} (\varphi, \varphi_\mu, \ldots , \varphi_{\mu_1, \ldots , \mu_{m-1}}; \pi^{\mu_1, \ldots , \mu_m}),
\end{equation}
by solving (\ref{genmom}). The Lorentz invariant Hamilton field equations then result to
\begin{eqnarray}
	\partial_{\mu} \varphi &=& \frac{\partial {\cal H} }{\partial \pi^\mu},  \quad  \,\,\,\,\,	\partial_\mu \varphi_\nu = \frac{\partial {\cal H} }{\partial \pi^{\mu \nu}},  \quad \ldots \quad  \, \partial_\nu  \varphi_{\mu_1 \dots \mu_{m-1}  }  =  \frac{\partial {\cal H} }{\partial \pi^{\nu \mu_1 \ldots \mu_{m-1}} } , \label{HFE1}\\   
	\partial_\mu \pi^\mu &=& -\frac{\partial {\cal H} }{\partial \varphi},  \quad  \partial_\nu \pi^{\mu \nu}= -\frac{\partial {\cal H} }{\partial \varphi_\mu},  \quad \ldots \,\,  \partial_\nu  \pi^{\mu_1 \dots \mu_{m-1} \nu }  = - \frac{\partial {\cal H} }{\partial \varphi^{\mu_1 \ldots \mu_{m-1}} } .  \label{HFE2}
\end{eqnarray}	
The generalisation of the above to relativistic scalar field theories involving multiple fields is straightforward. For our purposes here we reduce the dependence on the $x$-derivatives, thus ending up with nonrelativistic scalar field theory.

\subsection{Nonrelativistic formulation}

For a nonrelativistic theory we reduce the set of values $\mu_i$ can take and consider Lagrangian densities that only involve a first order derivative in $x$ and time-derivatives up to order $m$ 
\begin{equation}
	{\cal L}(\varphi, \varphi_x , \varphi_t \ldots , \varphi_{mt}).   \label{Lagnonrel}
\end{equation}
In the defining relations of the momenta we take $\mu_i=t$ for $i=1,\ldots,m$ and do not associate any canonical momenta to derivatives with respect to $x$. Thus (\ref{genmom}) becomes 
\begin{equation}
	\pi^{mt} := \frac{\partial  {\cal L} }{\partial \varphi_{mt}  }, \qquad
	\pi^{nt} := \frac{\partial  {\cal L} }{\partial \varphi_{nt}  }   - \partial_{t} 	\pi^{(n+1)t },   \quad n=1,\ldots, m-1.  \label{genmom2}
\end{equation}   
Replacing iteratively the generalized momenta $\pi^{(n+1)t }$ in the second relation in (\ref{genmom2}) we obtain 
\begin{equation}
	\pi_\ell :=	\pi^{\ell t} =  \sum_{k=\ell}^m (-1)^{k-\ell} \partial^{k-\ell}_t \left(\frac{\partial {\cal L}}{ \partial \varphi_{kt} } \right), \qquad \ell = 1, \ldots , m,  \label{pirel}
\end{equation}   
which corresponds to a formula that may be found in \cite{weldon03quant} after (2.11) in there. The equal-time canonical Poisson bracket relations between the canonical momentum fields $\pi_i$ and the canonical coordinate fields $\Phi_i$ are 
\begin{equation}
	\left\{   \Phi_i (x)  , \pi_j (x')      \right\}  = \delta_{ij}  \delta(x- x'), \qquad   \text{with}   \,\,\,       \Phi_i  := \varphi_{(i-1)t}  , \quad i,j =1, \ldots , m.   \label{canrel}
\end{equation}   
The Legendre transformation leading to the Hamiltonian density is then entirely expressed in terms of these fields
\begin{equation}
	{\cal H}=  \sum_{k=1}^{m-1}  \pi_k \Phi_{k+1} + \pi_m \hat{ \varphi}_{mt} - {\cal L}(\Phi_1,\ldots, \Phi_m,\hat{ \varphi}_{mt}  ) , \label{Ham22}
\end{equation}
where $\hat{ \varphi}_{mt}  $ means that we replace
\begin{equation}
	\varphi_{mt} \rightarrow \hat{ \varphi}_{mt} (\Phi_{1} , \ldots,\Phi_{m}; \pi_m);
\end{equation}
by solving (\ref{genmom2}). Since $\varphi_{mt}$ is not a canonical coordinate field one has achieved in this way that ${\cal H}$ only depends on canonical coordinate and momentum fields. The time-evolution of the canonical fields is then given by
\begin{equation}
	\left( \Phi_i  \right)_t =  \frac{\delta {\cal H}}{ \delta \pi_i}	= \sum_{n=0}^\infty (-1)^n \frac{\partial {\cal H}}{\partial  \left[ (\pi_i)_{nx}  \right]}, \qquad
	\left( \pi_i  \right)_t = - \frac{\delta {\cal H}}{ \delta \Phi_i} = \sum_{n=0}^\infty (-1)^n \frac{\partial {\cal H}}{\partial \left[ (\Phi_i)_{nx} \right] }  .   \label{timeevnon}
\end{equation}
Notice that ${\cal H}$ in (\ref{Ham22}) is not obtained as a direct special case from (\ref{Hamildens}), as a term of the form $\pi^x \varphi_x$ is not included, since we have not identified $\pi^x $ as a canonical momentum.

\subsection{Higher time-derivative theories in disguise as multi-field theories}

It is well known, see e.g. \cite{Nutku}, that the canonical Hamiltonian for the standard KdV equation can be derived from a Legendre transformed Lagrangian involving two scalar fields, instead of one, upon the implementation of Dirac constraints. For the space-time rotated version this implies that the higher time-derivatives are hidden in some extra fields, so that one can simply use the standard variational scheme with no need to invoke the conceptually more involved Ostrogradsky method at the cost of more algebraic complexity. In this section we show in more generality how to systematically reformulate HTDT as multi-field theories. Our starting point is a higher time-derivative Lagrangian of the form (\ref{Lagnonrel}) and the aim is to construct a transformation $\Gamma$ that converts it into a multi-field Lagrangian ${\cal L}'$ involving at most first order derivatives in time 
\begin{equation}
 	{\cal L}(\varphi, \varphi_x , \varphi_t \ldots , \varphi_{mt})  \xrightarrow{\Gamma}    {\cal L}' \left( \phi_1, (\phi_1)_x, (\phi_1)_t,  \phi_2, (\phi_2)_t, \ldots ,  \phi_{m'}, (\phi_{m'})_t \right) ,
\end{equation}
where $m' := [m/2 +1]$ with $[x]$, being the greatest integer function returning the greatest integer less than or equal to $x$. We demand that ${\cal L} \equiv {\cal L}'$ up to surface terms in the time integrations so that the resulting Euler-Lagrange equations from both are identical. For ${\cal L}'$ they are simply computed in the conventional manner as
\begin{equation}
         \sum_{n=0}^{m'}  \left\{  \frac{\partial {\cal L}' }{\partial \phi_n  } - \partial_t \left[ \frac{\partial {\cal L}'}{ \partial (\phi_n)_t }   \right] \right\}   =    \partial_x \left[ \frac{\partial {\cal L}'}{ \partial (\phi_n)_x} \right] . \label{higherEL}
\end{equation}
The conversion $\Gamma$ is achieved by relating the canonical momenta $\pi_\ell$ obtained from ${\cal L}$ by means of (\ref{pirel}) to $\pi'_\ell$ computed from ${\cal L}'$ as   
\begin{equation}
	\pi_\ell = (-1)^\ell    \frac{  \partial^{\ell-1} \pi'_\ell}{ \partial t^{\ell -1} } ,  \qquad \text{with}  \,\,  \pi'_\ell =  \frac{\partial {\cal L}' }{ \partial \phi_{2 \ell -1} }, \quad \ell= 1, \ldots, m' .  \label{constdes}
\end{equation}
Having computed all canonical momenta $\pi_\ell$, we solve these constraints by integrating out these equations to determine ${\cal L}'$ 
\begin{equation}
	{\cal L}' =  \!\! \int \left[  (-1)^\ell    \int  \pi_\ell dt^{\ell-1} \right]   d \varphi_{ (2 \ell -1)t }   + f_\ell \left(  \varphi, \varphi_x , \varphi_t \ldots   \check{\varphi}_{ (2 \ell -1)t } \ldots  , \varphi_{(m+1)t}   \right), \,\, \ell= 1,\ldots, m' ,  \label{intconstdes}
\end{equation}
where $ \check{\varphi}_{ (2 \ell -1)t }$ indicates that the dependence on these field is missing in $f_\ell $, as they result as integration functions from the integrations with respect to these field. Subsequently the unknown functions $f_\ell$ can be determined by comparing the $m' $ equations. As we will see below the time integrations over the canonical momenta $ \pi_\ell $ are often trivially performed.

\subsection{Higher time-derivative Hamiltonians from space-time rotated higher charges}

Our intention is to apply the above schemes not only to the standard space-time rotated Hamiltonians, but also to rotated versions of higher charges $Q_n$ interpreted as Hamiltonians. Here we briefly summarise how we construct our models. We start with a Hamiltonian density $ {\cal H}$ associated to an integrable system so that we can construct higher charge densities ${\cal Q}_n$. Subsequently, we interpret these charge densities as Hamiltonian densities, and use an inverse Legendre transform to construct their associated Lagrangian densities ${\cal L}_n$. We then exchange space and time leading to a ``rotated" Lagrangian $ {\cal L}^r_n$, which, given the nature of the models we started with, will inevitably involve higher order derivatives in time. Finally we employ Ostrogradsky's generalised scheme to derive the corresponding Hamiltonian densities ${\cal H}^r_n $. Our construction scheme is summarised as
\begin{equation}
	{\cal H}	 \xrightarrow{\text{integrability}} {\cal Q}_n \xrightarrow{\text{inverse Legendre transform}}  {\cal L}_n
	\xrightarrow{ x \leftrightarrow t}  {\cal L}^r_n  \xrightarrow{\text{Ostrogradsky scheme}} {\cal H}^r_n 
	\label{constscheme}
\end{equation}	
The conserved charges $Q_n = \int {\cal Q}_n dx$ involving the densities ${\cal Q}_n$ obey conservation laws of the form
\begin{equation}
	\left({\cal Q}_n\right)_t = 	\left(\chi_n\right)_x , \qquad \Rightarrow  \qquad  \frac{d Q_n}{dt} =  \int \left({\cal Q}_n\right)_t dx = \int \left(\chi_n\right)_x dx =0 ,  \label{240}
\end{equation}
where we assume that the flux $\chi_n $ vanishes or exactly cancels at the boundaries. Note that while the Euler-Lagrange equations maintain their form under an exchange of $x$ and $t$, that is not the case for Hamilton's field equations. The equations resulting from ${\cal H}^r_n $ are therefore different from those obtained from simply rotating ${\cal Q}_n$ and will lead to a new set of equations of motion.

Next we apply the schemes from above to various integrable field theoretical systems. 

\section{Canonical higher time-derivative Hamiltonians}
In this section, we derive canonical Hamiltonians, their canonical variables including their mutual Poisson brackets for a family of HTDT corresponding to rotated versions of general mKdV Hamiltonians with a space-time exchange and those resulting from higher charges as outlined in section 2.4. We compare the different schemes from sections 2.1, 2.2 and 2.3, one presenting a Lorentz-invariant version leading to Hamiltonians that are in general not preserved over time, one leading to a nonrelativistic version with conserved Hamiltonians and an equivalent version in terms of multiple fields with time-derivatives at most of order one.

\subsection{Standard Hamiltonian for modified KdV systems, generic $n$} 

In order to provide a point of reference, we start with the standard Lagrangian for the particle representation of the KdV system \cite{Graham} in a slightly generalised form that includes modified KdV systems. It is well known \cite{Nutku}, that the canonical Hamiltonian for the standard KdV equation can be obtained through a Legendre transformation of a Lagrangian involving two scalar fields, $\phi$ and $\psi$, along with their first-order derivatives, after the implementation of Dirac constraints. The Lagrangian for the mKdV system reads
\begin{equation}
	{\cal L} = \frac{1}{2} \phi_t \phi_x + \phi_x \psi_x + \phi_x^n + \frac{1}{2} \psi^2,     \qquad n \in \mathbb{N}, \label{LagKdV}
\end{equation} 
where $n=3$ and $n=4$ are special as they correspond to the integrable standard KdV system and standard modified KdV system, respectively. For simplicity we refer here by modified KdV systems to all cases with integer values $n \neq 3$. The systems for $n>4$ are not integrable, see e.g.  \cite{Miura}. 
The Euler-Lagrange equations derived from $	{\cal L} $ become
\begin{eqnarray}
	&&	\frac{ \partial {\cal L} }{\partial \phi} - \partial_\mu \left(   \frac{ \partial {\cal L} }{\partial_\mu \phi}   \right) =0, \quad \Rightarrow \quad \phi_{xt} + n(n-1) \phi_x^{n-2} \phi_{xx} + \psi_{xx} =0, \label{EL1}\\
	&&\frac{ \partial {\cal L} }{\partial \psi} - \partial_\mu \left(   \frac{ \partial {\cal L} }{\partial_\mu \psi}   \right) =0, \quad \Rightarrow \quad \psi=\phi_{xx}.
\end{eqnarray} 
Thus, one sees that the field $\psi$ was just concealing the higher $x$ derivatives in the $\phi$-field. When introducing the field $u:=\phi_x$, equations (\ref{EL1}) acquire the form of the modified KdV equations
\begin{equation}
	u_t + n(n-1) u^{n-2} u_x + u_{xxx} =0 .   \label{stanKdV}
\end{equation} 
As was pointed out in \cite{Nutku}, the Lagrangian ${\cal L}$ is degenerate, since the equations for the momentum fields $\pi_\psi = \partial {\cal L} / \partial \psi_t  = 0$ and $\pi_\phi = \partial {\cal L} / \partial \phi_t  = \phi_x/2$ can not be solved for the velocity fields, and therefore one can not convert from velocity phase space to the momentum phase space in a straightforward manner. However, by properly including Dirac constraints the canonical Hamiltonian with the correct Poisson bracket structure can be constructed, see  \cite{Nutku}. The unconstrained part of this Hamiltonian density is obtained from a Legendre transform and reads    
\begin{equation}
	{\cal H}_0 = \pi_\phi \psi_t + \pi_\phi \phi_t - {\cal L} = - \phi_x^3 - \phi_x \psi_x -\frac{1}{2} \psi^2 .
\end{equation} 
In the Hamiltonian we convert terms by integration by parts and a subsequent dropping of  surface terms in the usual way by assuming that all fields and their $x$-derivatives vanish at the boundaries. In terms of the KdV-fields we then obtain
\begin{equation}
	H_0 = \int  {\cal H}_0  dx = \int \left( \frac{1}{2} u_x^2 -u^n \right) dx.  \label{HamH0}
\end{equation} 
Here we are mainly interested in the energies of particular solution and since the additional term $H_1$, required to derive the correct Poisson bracket structure, does not contribute to them we will not report it here, but simply refer to \cite{Nutku} for its explicit expression.

\subsection{Rotated standard Hamiltonian for modified KdV systems, generic $n$} 

A space-time rotated, one scalar field, partially integrated surface term adjusted version of the Lagrangian (\ref{LagKdV}) is
\begin{equation}
	{\cal L}^r = \frac{1}{2} \varphi_t \varphi_x +  \varphi_t^n - \frac{1}{2} \varphi_{tt}^2,    \qquad n \in \mathbb{N}. \label{L235}
\end{equation}
At first we derive the Lorentz invariant version of Hamilton's equations by the scheme as outlined in section 2.1. 
Using the defining relations for the generalised momenta (\ref{genmom}) with $m=2$, we compute them to
\begin{equation}
	\pi^x = \frac{1}{2} \varphi_t, \quad 	\pi^t = \frac{1}{2} \varphi_x + \varphi_{ttt} + n \varphi_t^{n-1}, \quad
	\pi^{tt}= - \varphi_{tt},  \quad  \pi^{xx}= \pi^{xt}= \pi^{tx}= 0 .    \label{canmom32}
\end{equation}
The Hamiltonian density according to (\ref{Hamildens}) then results to
\begin{eqnarray}
	{\cal H}^r &=& \pi^x \varphi_x +\pi^t \varphi_t  + \pi^{tt} \hat{\varphi}_{tt} - 	{\cal L}^r(\varphi , \varphi_t, \varphi_x, \hat{\varphi}_{tt} )  \\
	&=& \pi^x \varphi_x +  \pi^t \varphi_t  - \frac{1}{2} \left(  \pi^{tt} \right)^2  - \frac{1}{2} \varphi_t \varphi_x -\varphi_t^3, 
\end{eqnarray}
where we identified $\hat{\varphi}_{tt}  =- \pi^{tt} $. Hamilton's field equations are then consistently computed according (\ref{HFE1}) and (\ref{HFE2})
\begin{eqnarray}
	\partial_\mu \varphi &=& \frac{{{\cal H}}^r }{ \partial \pi^\mu}   \quad \Leftrightarrow   \quad    \varphi_t = \varphi_t , \,\,  \varphi_x = \varphi_x, \label{e240} \\ 
	\partial_\mu \varphi_\nu &=& \frac{{{\cal H}}^r }{ \partial \pi^{\mu \nu} }    \quad \Leftrightarrow   \quad   \varphi_{tt}  = - \pi^{tt},  \\
	\partial_\mu \pi^\mu&=& - \frac{{{\cal H}}^r }{\partial \varphi}   \,\, \Leftrightarrow   \,\,
	\partial_t \pi^t + 	\partial_x \pi^x =0,  \,\, \Leftrightarrow   \,\, n(n-1) \varphi_t^{n-2} \varphi_{tt} + \frac{1}{2} \varphi_{xt} + \varphi_{tttt} + \frac{1}{2} \varphi_{xt}=0, \qquad \,\,\,
	\\
	\partial_\nu \pi^{\mu \nu}&=& - \frac{{{\cal H}}^r }{ \partial \varphi_\mu}   \quad \Leftrightarrow   \quad 
	\partial_t \pi^{tt} = -   \frac{{{\cal H}}^r }{ \partial \varphi_t}    \quad \Leftrightarrow   \quad 
	-\varphi_{ttt} = - \pi^t  + \frac{1}{2} \varphi_x + n \varphi_t^{n-1} .   \label{e243}
\end{eqnarray}	

While the Hamiltonian density $	{{\cal H}}^r $ yields a consistent set of Hamilton's equations in the relativistic framework its corresponding Hamiltonian $ { H}^r = \int {{\cal H}}^r  dx$ is in general not a conserved quantity
\begin{equation}
	\frac{ d{ H}^r}{dt} = \int \frac{ d { {\cal H}}^r}{dt}  dx  =  \int \frac{ d ( \pi^x \varphi_x )}{dt}  dx  
	+  \int \frac{ d  {\cal H}'^r}{dt}  dx =   \int \frac{ d ( \pi^x \varphi_x) }{dt}  dx \neq 0 .
\end{equation}
We will present the precise form of the conserved Hamiltonian ${\cal H}'^r$ below, which is in fact the nonrelativistic version of $ H^r$, which we should be using as the modified KdV systems are not Lorentz invariant.

We will demonstrate this scheme for the equivalent Lagrangian ${\cal L}'^r $ obtained from ${\cal L}^r$ by initially hiding the higher derivative fields as explained in general terms in section 2.3.  With the canonical momenta computed as in (\ref{canmom32}) the integrated constraints (\ref{constdes}) yield with (\ref{intconstdes})
\begin{eqnarray}
	  {\cal L}'^r  &=& \int \pi^t d \varphi_t = \frac{1}{2} \varphi_x  \varphi_t +  \varphi_t^n +  \varphi_t  \varphi_{ttt} + f_1(\varphi ,\varphi_{tt} ,\varphi_{ttt}) ,  \label{11221}  \\
	   {\cal L}'^r  &=& -\int \left[ \int \pi^{tt} dt  \right] d \varphi_{ttt} = \varphi_{tt} \varphi_{ttt} + f_2(\varphi ,\varphi_{t}  \varphi_{tt}) .   \label{11222} 
\end{eqnarray}
The comparison of (\ref{11221}) and (\ref{11222}) then gives
\begin{equation}
	{\cal L}'^r  =   \frac{1}{2} \varphi_x  \varphi_t +  \varphi_t^n +  \varphi_t  \varphi_{ttt} + g(\varphi, \varphi_{tt}), 
\end{equation}
where $g(\varphi, \varphi_{tt})$ is the last unknown function. Equating ${\cal L}^r  $ and ${\cal L}'^r  $ we identify at first $g(\varphi, \varphi_{tt}) = -  \varphi_t  \varphi_{ttt} - 1/2  \varphi_{tt}^2 $, but since $g$ does not depend on  $\varphi_{ttt}$ we must convert the first term with an integration by parts, so that  $g(\varphi, \varphi_{tt}) =  1/2  \varphi_{tt}^2$ when surface terms are dropped. Finally, we obtain
\begin{equation}
	{\cal L}'^r  =   \frac{1}{2} \varphi_x  \varphi_t +  \varphi_t^n +  \varphi_t  \varphi_{ttt} + \frac{1}{2}  \varphi_{tt}^2 ,
\end{equation}
which differs from ${\cal L}^r $ just by surface terms as anticipated. Introducing now the new fields $\phi := \varphi$ and $\psi := \varphi_{tt}$, the Lagrangian ${\cal L}'^r  $ acquires the form
\begin{equation}
	{\cal L}'^r = \frac{1}{2} \phi_t \phi_x + \phi_t \psi_t + \phi_t^n + \frac{1}{2} \psi^2,    \qquad n \in \mathbb{N},   \label{Lrot}
\end{equation}
so that indeed all higher order time-derivatives have been hidden. When rotating back by swapping $x$ and $t$ we recover the version (\ref{LagKdV}).

The Euler-Lagrange equations are now obtained from ${\cal L}'^r$ by using (\ref{higherEL}) to 
\begin{equation}
	\phi_{xt} + n(n-1) \phi_t^{n-2} \phi_{tt} + \psi_{tt} =0, \qquad \text{and} \qquad \psi=\phi_{tt}. \label{mKdVphi}
\end{equation}
With $u:= \phi_t$ the first equations converts into the rotated mKdV equations
\begin{equation}
	u_x + n(n-1) u^{n-2} u_t + u_{ttt} =0 .  \label{rotKdV}
\end{equation} 
Unlike the original unrotated Lagrangian, the rotated Lagrangians ${\cal L}^r$ and ${\cal L}'^r$ are not degenerate. The canonical momenta for ${\cal L}'^r$ are computed directly to 
\begin{equation}
	\pi_\psi = \frac{\partial {\cal L}'^r}{\partial \psi_t} = \phi_t, \qquad \pi_\phi = \frac{\partial {\cal L}'^r}{\partial \phi_t} = \frac{1}{2} \phi_x + n \phi_t^{n-1} + \psi_t,
\end{equation}
which can be inverted for the velocity fields  
\begin{equation}
	\phi_t = 	\pi_\psi , \qquad \psi_t = \pi_\phi - \frac{1}{2} \phi_x - n \pi_\psi^{n-1} .
\end{equation}
The Legendre transformed Lagrangian ${\cal L}'^r$ then yields the Hamiltonian density 
\begin{equation}
	{\cal H}'^r = \pi_\psi \psi_t + \pi_\phi \phi_t - {\cal L}'^r = \pi_\phi \pi_\psi- \frac{1}{2}   \pi_\psi  \phi_x - \pi_\psi^n - \frac{1}{2} \psi^2  .
\end{equation} 
This is indeed the canonical Hamiltonian with the correct underlying canonical Poisson bracket structure. As a crucial difference compared to the nonrotated version we did not require any Dirac constraints in its derivation. We assume the standard equal-time canonical Poisson bracket relations for the canonical fields
\begin{equation}
	\left\{  \psi(x) , \pi_\psi(x')      \right\}     =  \delta \left( x- x'  \right), \qquad
	 \left\{  \phi(x) , \pi_\phi(x')      \right\}  =  \delta \left( x- x'  \right),
\end{equation}
with all other combinations of $\psi,\phi,\pi_\psi , \pi_\phi$ vanishing. When translating the vanishing relations to the $\phi$-fields and their time-derivatives we obtain 
\begin{equation}
	\left\{  \phi_{tt}(x) , \phi_{ttt}(x')      \right\}     = -n(n-1) \phi_t^{n-2} \partial_x \delta \left( x- x'  \right), \quad
	\left\{  \phi_{ttt}(x) , \phi_{ttt}(x')      \right\}     = - \partial_x \delta \left( x- x'  \right),
\end{equation}
and
\begin{equation}
     	\left\{  \phi(x) , \phi(x')      \right\}  =  	\left\{  \phi(x) , \phi_t(x')      \right\} = 	\left\{  \phi(x) , \phi_{tt}(x')    
     	  \right\}   = 	\left\{  \phi_t(x) , \phi_t(x')      \right\} = 	\left\{  \phi_t(x) , \phi_{ttt}(x')      \right\} =0 .
\end{equation}
With the help of these relations we compute Hamilton's equations of motion
\begin{eqnarray}
	\phi_t &=& \frac{ \partial { H}'^r }{\partial \pi_\phi} = \left\{  \phi, { H}'^r   \right\}  = \pi_\psi,  \\ 
	\psi_t &=& \frac{ \partial { H}'^r }{\partial \pi_\psi} = \left\{  \psi, { H}'^r   \right\}  = \pi_\phi - \frac{1}{2} \phi_x - n \pi_\psi^{n-1},  \\ 
	\left(  \pi_\psi   \right)_t &=&  -\frac{ \partial { H}'^r }{\partial \psi} = \left\{   \pi_\psi , { H}'^r   \right\}  = \psi = \phi_{tt} , \\
	\left(  \pi_\phi   \right)_t &=&  -\frac{ \partial { H}'^r }{\partial \phi} = \left\{   \pi_\phi , { H}'^r   \right\}  = - \frac{1}{2} 	\left(  \pi_\psi   \right)_x , \label{217}
\end{eqnarray} 
where (\ref{217}) is the first equation of (\ref{mKdVphi}). We can express the Hamiltonian density in terms of the KdV-field $u(x,t)$ as
\begin{equation}
	{\cal H}'^r =  (n-1) u^n + u u_{tt} -\frac{1}{2} u_t^2,  \label{Haminu}
\end{equation}
which is a conserved quantity as is easily verified when using the equation of motion (\ref{rotKdV})
\begin{equation}
	\frac{d H'^r}{dt} = \int \frac{d {\cal H}'^r}{dt}  dx = \int u \left[u_{ttt} +n(n-1)u^{(n-2)} u_t \right]dx = - \frac{1}{2}\int (u^2)_x  dx =0 ,
\end{equation}
and when dropping surface terms.
The nonvanishing equal-time Poisson brackets for the KdV fields are obtained by direct convertion of the above
\begin{eqnarray}
	\left\{  u(x) , u_t(x')      \right\}     &=& -\delta \left( x- x'  \right), \\
	\left\{  u_{tt}(x) , u_{tt}(x')      \right\}     &=& - \partial_x \delta \left( x- x'  \right), \\
		\left\{  u_{t}(x) , u_{tt}(x')      \right\}    & =&  -n(n-1) u^{n-2} \partial_x \delta \left( x- x'  \right),
\end{eqnarray} 
such that
\begin{equation}
	u_t    = \left\{  u , H'^r \right\} .
\end{equation} 
Thus, $H'_r$ is a conserved quantity interpreted in the usual fashion as energy and also as a Hamiltonian since it generates the evolution in time.  Notice that $H_r$ does not admit such an interpretation.

Solutions $u_4$ and $u_3$ to the equation of motion (\ref{rotKdV}) for $n=4$ and $n=3$, respectively, are related by means of the rotated Miura transformation 
\begin{equation}
	u_3    = 2 u_4^2 + i \sqrt{2} (u_4)_t,   \label{Miura}
\end{equation} 
offering the possibility to construct a complex solution $u_3$ from a real solution $u_4$ or possibly a real solution  $u_3$ from a complex solution $u_4$.  

\subsection{xt-rotated first higher charge Hamiltonians for the KdV system, $n=3$} 
Next we discuss a HTDT obtained from a rotated higher charge following the scheme outlined in (\ref{constscheme}).
Starting with the  KdV system, the next highest charge density satisfying (\ref{240}) beyond the Hamiltonian  reads
\begin{equation}
	{\cal Q}_3 =  \frac{5 }{3} u u_x^2 -\frac{5 }{6} u^4 -\frac{1}{6} u_{xx}^2,
\end{equation} 
with corresponding flux
\begin{equation}
	\chi_3 = \frac{1}{6} \left[ 24 u^5-u_{xxx}^2  +10 \left(2 u^3+u_x^2\right) u_{ xx }-90 u^2 u_x^2+2 u_{4 x}	u_{xx }-20 u u_x u_{xxx }+16 u u_{ xx}^2\right] .
\end{equation} 
Identifying $u:= \varphi_x $, we obtain the associated Lagrangian by an inverse Legendre transform as
\begin{equation}
	{\cal L}_3 =  \frac{1}{2} \varphi _t \varphi _x-\frac{5}{3} \varphi _x \varphi _{xx }^2+\frac{5 }{6} \varphi _x^4+\frac{\varphi
		_{xxx }^2}{6},
\end{equation} 
Following the scheme in (\ref{constscheme}), the rotated version 
\begin{equation}
	{\cal L}_3^r =  \frac{1}{2} \varphi _t \varphi _x-\frac{5}{3} \varphi _t \varphi _{tt}^2+\frac{5 }{6} \varphi _t^4+\frac{\varphi
		_{ttt }^2}{6},
\end{equation} 
is used next as the starting point of the Ostrogradsky procedure.  The generalised momentum and coordinate fields are computed from (\ref{genmom2}) and (\ref{canrel}) to
\begin{eqnarray}
	\pi_1 &=& \frac{10}{3} \left(   \varphi _t \varphi _{ttt} + \varphi _t^3 \right)
	    +\frac{5 \varphi
		_{tt}^2}{3} +\frac{\varphi _{5 t}}{3} +\frac{\varphi _x}{2},  \,\, \, \pi_2 = -\frac{10}{3} \varphi _t \varphi _{tt}-\frac{\varphi _{4 t}}{3}, \,\,\, \pi_3= \frac{\varphi _{ttt}}{3} , \,\,\,\, \label{mom234} \\
		\Phi_1 &=& \varphi,  \,\,\,   \Phi_2 = \varphi_t,  \,\,\, \Phi_3 = \varphi_{tt} .
\end{eqnarray} 
Also this system is not degenerate so that the Hamiltonian density is obtained directly from (\ref{Ham22})
\begin{eqnarray}
	{\cal H }'^{r}_{3} &=&  \pi_1 \Phi _2+ \pi_2 \Phi _3+\frac{3 }{2} \pi_3^2   -\frac{1}{2}
	 \Phi _2  \left( \Phi _3 \right)^2+ \frac{5}{3} \Phi_2 \Phi_3^2-\frac{5 }{6} \Phi_2^4   ,\\
	&=&  -\frac{u_t u_{ ttt }}{3}+\frac{1}{3} u u_{4 t}+\frac{10 u^2
		u_{tt}}{3}+\frac{u_{tt}^2}{6}+\frac{5 u^4}{2} ,   \label{336z}
\end{eqnarray}	
with $u:= \varphi_t$.

Hamilton's equation of motion lead again to many trivially satisfied identities with the equation of motion resulting to
\begin{eqnarray}
	\left( \pi_1 \right)_t= - \frac{{\cal H }'^r_3 }{\partial \varphi} \,\, &\Leftrightarrow& \varphi_{xt}+10 \varphi _t^2 \varphi _{tt }+\frac{10}{3}\varphi _t \varphi _{4 t} +\frac{1}{3} \varphi _{6 t}+\frac{20 }{3} \varphi
	_{tt } \varphi _{ ttt  },\\
	&\Leftrightarrow& u_x +10 u^2 u_t+\frac{10   }{3} u u_{ ttt } +\frac{1}{3} u_{5 t}+ \frac{20}{3}  u_t u_{ tt }, \label{24811}
\end{eqnarray}	
 We observe that ${\cal H }'^{r}_{3}$  is a conserved quantity 
\begin{eqnarray}
	\frac{d H'^r_3}{dt} &=& \int \frac{d {\cal H}'^r_3}{dt}  dx = \int u \left[  10 u^2 u_t+\frac{10   }{3} u u_{ ttt } +\frac{1}{3} u_{5 t}+ \frac{20}{3}  u_t u_{ tt }       \right]dx \\
	&=& - \frac{1}{2}\int (u^2)_x  dx =0 ,
\end{eqnarray}
up to the validity of the equation of motion (\ref{24811}) and the vanishing of surface terms.

\subsubsection{Multi-field theory}
Next we demonstrate that one may also re-express the higher charge Lagrangian ${\cal L}'^r_3 $ and Hamiltonian ${\cal H}'^r_3$  densities in terms of new fields that hide the higher time-derivatives. Solving the constraints (\ref{constdes}) using the canonical momenta (\ref{mom234}) we obtain the equivalent multi-field Lagrangian
	\begin{eqnarray}			
	\mathcal{L}'^r &=&\frac{5}{6}\varphi_t^4+\frac{5}{3}\varphi_{2t}^2\varphi_t+\frac{1}{2}\varphi_x\varphi_t
	+\frac{5}{3}\varphi_t^2\varphi_{3t}+\frac{1}{3}\varphi_{5t}\varphi_t+\frac{1}{6}\varphi_{3t}^2  +\frac{1}{3}   \varphi_{2t} \varphi_{4t},		
	\label{nearlyfixed2}		\\
	&=&  \frac{5}{6}\left( \phi_1 \right)_t^4+\frac{5}{3}\phi_2^2\left(\phi_1\right)_t 	+\frac{1}{2}\left(\phi_1\right)_x\left(\phi_1\right)_t+\frac{5}{3}\left(\phi_1\right)_t^2\left(\phi_2\right)_t +\frac{1}{3}\left(\phi_3\right)_t\left(\phi_1\right)_t+\frac{1}{6}\left(\phi_1\right)_t^2+\frac{1}{3}  \phi_1 \phi_2 .
\notag
\end{eqnarray}
where $\phi_1 := \varphi$,  $\phi_2 := \varphi_{2t}$ and  $\phi_3 := \varphi_{4t}$. Computing the corresponding canonical momenta gives 
\begin{eqnarray}
	\pi'_{1}&=&\frac{\partial \mathcal{L}'^r}{\partial \left(\phi_1\right)_t}=\frac{10}{3}\left(\phi_1\right)_t^3+\frac{5}{3}\phi_2^2+\frac{1}{2}\left(\phi_1\right)_x+\frac{10}{3}\left(\phi_1\right)_t\left(\phi_2\right)_t+\frac{1}{3}\left(\phi_3\right)_t , \label{expi1}\\
	\pi'_{2}&=&\frac{\partial \mathcal{L}'^r}{\partial \left(\phi_2\right)_t}=\frac{5}{3}\left(\phi_1\right)_t^2+\frac{1}{3}\left(\phi_2\right)_t ,\\
	\pi'_{3}&=&\frac{\partial \mathcal{L}'^r}{\partial \left(\phi_3\right)_t }=\frac{1}{3}\left(\phi_1\right)_t , \label{expi3}
\end{eqnarray}	
so that the Legendre transformed $\mathcal{L}'^r $ yields the Hamiltonian
\begin{eqnarray}
\mathcal{H}'^r &=&  3 \pi'_{3}  \pi'_{1}+270 {\pi'}_{\! \!3}^{4} -5\left(\phi_2\right)^2 {\pi'}_{ \!\! 3}-\frac{3}{2} {\pi'}_{\!\! 3} \left( \phi_1\right)_x-45 {\pi'}_{\!\!3}^2 {\pi'}_{\!\! 2}+\frac{3}{2} {\pi'}_{\!\!2}^2-\frac{1}{3} \phi_2 \phi_3 ,   \label{hhh11} \\
	&=&  \frac{5}{2}\left(\phi_1\right)_t^4+\frac{10}{3}\left(\phi_1\right)_t^2\left(\phi_2\right)_t+\frac{1}{6}\left(\phi_2\right)_t^2+\frac{1}{3}\left(\phi_1\right)_t\left(\phi_3\right)_t  -\frac{1}{3} \phi_2 \phi_3  . \label{hhh}
\end{eqnarray}	
With the identification $\varphi_t = u$ the expression (\ref{hhh}) is identical to (\ref{336z}) previously derived. The version (\ref{hhh11}) is tailored to generate the equations of motion. We find 
	\begin{eqnarray}
		\left(\phi_1\right)_t&=& \int \{\phi(x),\mathcal{H}(y)\}dy=3{\pi'}_{\!\! 3}
		\label{Q3phidotrot}\\
		\left(\phi_2\right)_t &=&\int \{\psi(x),\mathcal{H}(y)\}dy=-45 {\pi'}_{\!\! 1}^{2} +3 {\pi'}_{\!\! 2}
		\label{Q3psidotrot}\\
			\left(\phi_3\right)_t&=&\int \{\theta(x),\mathcal{H}(y)\}dy=3 {\pi'}_{\!\! 1} +1080 {\pi'}_{\!\! 3}^{3}  -5 \phi_1 ^2-\frac{3}{2}\left(\phi_0\right)_x-90{\pi'}_{\!\! 2} {\pi'}_{\!\! 3}
		\label{Q3thetadotrot} \\
		\left( {\pi'}_{\!\! 1} \right)_t &=&\int\{\pi_{\phi}(x),\mathcal{H}(y)\}dy=-\frac{3}{2}( {\pi'}_{\!\! 3}    )_x
		\label{Q3piphidotrot}\\
		\left( {\pi'}_{\!\! 2} \right)_t &=&\int\{\pi_{1}(x),\mathcal{H}(y)\}dy=10 \phi_1{\pi'}_{\!\! 3}   +\frac{1}{3} \phi_2 
		\label{Q3pipsidot}\\
			\left( {\pi'}_{\!\! 3} \right)_t &=&\frac{1}{3} \phi_1  
		\label{Q3pithetadotrot}
		\end{eqnarray}	
Using the explicit expression for the canonical momenta (\ref{expi1})-(\ref{expi3}) we see that all equations are trivially satisfied, except (\ref{Q3piphidotrot}) which corresponds to the equation of motion. Thus one can eliminate all time-derivatives beyond order one by introducing new fields in the described manner and simply use the standard variational principle, avoiding the use Ostrogradsky's method altogether. However, the conversion (\ref{intconstdes}) becomes increasingly complicated as we have checked for the next highest charge not reported here.

\subsubsection{Integrability from Painlev\'e test }
Having obtained a new version of a conserved  Hamiltonian it is unclear whether the system is integrable or not. In order to settle this question we carry out a Painlev\'e test \cite{Painor,ARS,Pain1,Pain2,Kruskal,Gramma} for the equation of motion in the form (\ref{24811}). Our starting point is a Painlev\'e expansion of the form
\begin{equation}
	u(x,t) = \sum_{k=0}^\infty \lambda_k(x,t) w(x,t)^{k+\alpha},  \label{Painexpan}
\end{equation}
where the functions $\lambda_k(x,t)$ are analytic and $-\alpha \in \mathbb{N}$ characterises the leading order singularity when $w(x,t) \rightarrow 0$. Substituting the expansion into the PDE under investigation the function $\lambda_k(x,t)$ may in principle be computed recursively. The Painlev\'e test consists of determining how many free parametric functions $\lambda_k(x,t)$, referred to as resonances, occur in this recursive process or whether the procedure would impose constraints on the functions $w(x,t)$. In case the number of resonances is greater or equal to the order of the differential equation the system is said to have passed the Painlev\'e test and is considered to be integrable. In reverse, if that is not the case the test is failed and one can deduce that the system is not integrable. We will not establish here the stronger Painlev\'e property, which consists of proving in addition that the series (\ref{Painexpan}) is convergent.

For our system we need to establish at first the order of the leading singularity. Substituting the first term from the expansion (\ref{Painexpan}) into (\ref{24811}) we can identify the leading order contributions from each of the terms $u_x \sim w^{\alpha-1}$,  $ u^2 u_t \sim w^{3 \alpha-1}$,   $ u u_{ ttt } \sim  w^{\alpha-3}$,  $ u_{5 t} \sim w^{\alpha-5}$, $ u_t u_{ tt } \sim w^{\alpha-2}$. Matching the powers of the second and fourth term, $ 3 \alpha -1 = \alpha -5 $, we find that $\alpha = -2$. Using this value in the expansion (\ref{Painexpan}), we read off the coefficients in the powers of $w$ and set them to zero
\begin{equation}
	w^{-7}: \quad    \lambda _0 w _t \left(\lambda _0+2 w_t^2\right) \left(\lambda _0+6 w _t^2\right) =0  . \label{w7}
\end{equation}	
For the solution $\lambda_0 = -2 w_t^2$  to (\ref{w7}) we find at the next orders
\begin{eqnarray}
	w^{-6}: &&  \,\,    w_t^5 \left(\lambda _1-2  w_{tt}\right) =0 \quad \Rightarrow  \quad   \lambda _1 = 2  w_{tt} ,    \label{261}  	 \\
w^{-5}:  && 
 \,\, w_t^3 \left(\lambda _1-6 w_{tt }\right) \left(1-2
 w_t \partial_t \right) \left(\lambda _1-2 w_{ tt }\right) =0, \quad \Rightarrow \quad   \lambda _2  \,
 \text{arbitrary}  , \label{w52}  \notag     \\
 w^{-4}:  && \,\, 
               \lambda_3 =  \frac{w_t \left[w_t \left(2 w_{ tt}
               	w_{ttt}-2 w_t \left(\lambda_2 \right)_t+w_{4 t}\right)\right]-w_{tt}^3}{2 w_t^4}  ,   \notag \\          	
w^{-3}:  && \,\,    \lambda_4 =   \frac{40 w_t w_{tt}^2 w_{ttt}+10 w_t^4 \left(\lambda _2\right)_{tt}-25 w_{4 t} w_t^2
	w_{tt}-20 w_{tt}^4+30 \lambda _2^2 w_t^4}{20 w_t^6}    \notag    \\
	&& \qquad \qquad+ \frac{10 \lambda _2 w_t^2 \left(4 w_t w_{ttt}-3 w_{tt}^2\right)+w_t^3 \left(6 w_{5 t}+3 w_x-10 \left(\lambda_2\right)_t w_{tt}\right)}{20 w_t^6}  , \notag \\
w^{-2}:  &&\,\,     \lambda _5  \,
\text{arbitrary}    ,  \notag     \\
w^{-1}:  && \,\,      \lambda _6  \,
\text{arbitrary}   ,   \notag       
\end{eqnarray}
\begin{eqnarray}
w^{0}:  && \,\,    \lambda _7 = \frac{1}{80 w_t^5} \left\{180 \lambda _4 w_{tt} w_{ttt}-40 \lambda _3^2 w_t w_{tt}-30 \lambda _5 w_t w_{tt}^2-80 \lambda _3 \lambda _4 w_t^3-10
\lambda _4 w_{4 t} w_t  \right.   \notag   \\
 &&  +21 \lambda _3 w_{5 t} + 30 \lambda _2^2 \left[   \left(\lambda_2 \right)_t +\lambda _3 w_t   \right]+45 w_{4 t} \left(\lambda_3 \right)_t-160 \lambda _6 w_t^3 w_{tt}-80 \lambda _5 w_t^2 w_{ttt}+3 \lambda _3
 w_x  \notag \\
  &&   +150 w_{tt}^2 \left(\lambda_4 \right)_t-120 w_t^2 w_{tt} 
  \left(\lambda_5 \right)_t+40 w_t w_{ttt}  \left(\lambda_4 \right)_t  -80 \lambda _3 w_t^2
      \left(\lambda_3 \right)_t   -80 w_t^4  \left(\lambda_6 \right)_t \notag \\
&& +  20 \left(\lambda_2 \right)_t \left[ 2 w_t \left(  \left(\lambda_3 \right)_t   -2 \lambda _4
      w_t\right)+  \left(\lambda_2 \right)_{tt}   +7 \lambda _3 w_{tt}\right]  +20 \lambda _3 w_t
     \left(\lambda_2 \right)_{tt}   -40 w_t^3 \left(\lambda_5 \right)_{tt} \notag  \\   
  &&      +   10 \lambda _2 \left[3 w_t \left[ \left(\lambda_3 \right)_{tt}   -2 \lambda _4 w_{tt}   -2 w_t \left(   \left(\lambda_4 \right)_{t}   +\lambda _5
  w_t\right)     \right]
     +15 w_{tt}
     \left(\lambda_3 \right)_{t}    + \left(\lambda_2 \right)_{ttt}   +13 \lambda _3 w_{ttt}\right]   \notag  \\   
      &&    \left.  + 30 w_{tt} \left(\lambda_3 \right)_{ttt}  +60 w_t w_{tt} \left(\lambda_4 \right)_{tt} +50 w_{ttt}  \left(\lambda_3 \right)_{tt}    +5 w_t \left(\lambda_3 \right)_{4t} + \left(\lambda_2 \right)_{5t} +3 \left(\lambda_2 \right)_{t}     \right\} ,\notag  \\   
w^{1}:  && \,\,     \lambda _8  \,  \notag 
\text{arbitrary}       .   
\end{eqnarray}
Thus, besides the fundamental singularity at $\lambda_{-1}$, we found the four additional free parameters $\lambda_2$, $\lambda_5$, $\lambda_6$ and $\lambda_8$. Thus, with five free parameters we match the order of the differential equation and conclude that the system is integrable. 

We compare this result of our explicit construction with the necessary condition which arises when we assume the $\lambda_r$ for some as yet unknown $r$ to be constant. We set them to some constant coefficient $\theta$. For the unknown power $r$ the constant $\theta$ of the term $w^{r + \alpha}$ becomes undetermined when this term cancels the leading order $w^{\alpha}$. Hence, substituting for this 
\begin{equation}
	\tilde{u}(x,t)   = \lambda_0(x,t) w(x,t)^\alpha +  \theta w(x,t)^{r+\alpha}   \label{nescond}
\end{equation} 
with $\lambda_0 = -2 w_t^2$ into (\ref{24811}), we obtain
\begin{equation}
	\frac{1}{3} \theta  (r-8) (r-6) (r-5) (r-2) (r+1) w_t^5 =0 ,
\end{equation}
which reproduces exactly the powers $r=2,5,6,8$found in the explicit construction above. For the other two choices for $\lambda_0=0$ and  $\lambda_0 = -6 w_t^2$, we find
\begin{eqnarray}
  	\frac{1}{3} \theta  (r-6) (r-5) (r-4) (r-3) (r-2) w_t^5 &=&0 ,\\
	  \frac{1}{3} \theta  (r-10) (r-8) (r-6) (r+1) (r+3) w_t^5 &=&0 ,
\end{eqnarray}
respectively. Once again we find five free parameters. We verified these values also in an explicit construction that we do not report here.

\subsection{xt-rotated first higher charge Hamiltonians for the KdV system, $n=4$} 
For the modified KdV system, the next highest charge density satisfying (\ref{240}) beyond the Hamiltonian  reads
\begin{equation}
	{\cal Q}_3 = \frac{10}{3} u^2 u_x^2-\frac{4 u^6}{3}-\frac{u_{xx}^2}{6},
\end{equation} 
with corresponding flux
\begin{equation}
	\chi_3 = 12 u^8+8 u^5 u_{xx}-60 u^4 u_x^2-\frac{20}{3} u^2 u_x u_{xxx}+\frac{16}{3} u^2
	u_{xx}^2+\frac{20}{3} u u_x^2 u_{xx}+\frac{1}{3} u_{4 x}
	u_{xx}+\frac{u_x^4}{3}-\frac{u_{xxx}^2}{6}.
\end{equation} 
Identifying again $u:= \varphi_x $, we obtain the associated Lagrangian by an inverse Legendre transform as
\begin{equation}
{\cal L}_3 =  \frac{1}{2} \varphi _t \varphi _x-\frac{10}{3} \varphi _x^2 \varphi _{xx}^2+\frac{4 }{3} \varphi _x^6 +\frac{\varphi
	_{xxx}^2}{6}.
\end{equation} 
Following the scheme in (\ref{constscheme}), the rotated version 
\begin{equation}
	{\cal L}_3^r =  \frac{1}{2} \varphi _x \varphi _t-\frac{10}{3} \varphi _t^2 \varphi _{tt}^2+\frac{4 }{3} \varphi _t^6 +\frac{\varphi
		_{ttt}^2}{6},
\end{equation} 
is used next as the starting point of the Ostrogradsky procedure. 
The generalised momentum and coordinate fields are computed from (\ref{genmom2}) and (\ref{canrel}) to
\begin{eqnarray}
	\pi_1 &=& \frac{20}{3}\left( \varphi _t \varphi _{tt}^2+\varphi _t^2 \varphi _{ttt} \right)+8 \varphi_t^5+\frac{\varphi _{5 t}}{3}+\frac{\varphi _x}{2},  \,\,\, \pi_2= -\frac{20}{3} \varphi _t^2 \varphi _{tt}-\frac{\varphi _{4 t}}{3}, \,\, \, \pi_3= \frac{\varphi _{ttt}}{3} ,  \qquad\\
	\Phi_1 &=& \varphi,  \,\,\,   \Phi_2 = \varphi_t,  \,\,\, \Phi_3 = \varphi_{tt} .
\end{eqnarray} 
Once again the system is not degenerate so that the Hamiltonian density is obtained directly from (\ref{Ham22})
\begin{eqnarray}
	{\cal H}'^r_3&=&  \pi_1 \Phi _2 + \pi_2 \Phi _3+\frac{3 }{2} \pi_3^2+\frac{10}{3}
	\varphi _t^2 \varphi _{tt}^2-\frac{1}{2} \varphi _t \varphi _x -\frac{4}{3}  \varphi _t^6,\\
	&=&   -\frac{u_t u_{ttt }}{3}+\frac{10}{3} u^2 u_t^2+\frac{1}{3} u u_{4 t}+\frac{20 u^3
		u_{tt }}{3}+\frac{u_{tt}^2}{6}+\frac{20 u^6}{3} ,   \label{371z}
\end{eqnarray}	
with $u:= \varphi_t$. Hamilton's equation of motion lead once more to many trivially satisfied identities, that we do not report, with the equation of motion resulting as
\begin{eqnarray}
	\left(  \pi_1 \right)_t = - \frac{{\cal H}^r_3 }{\partial \varphi} \,\, &\Leftrightarrow& \varphi_{xt} +\frac{80}{3} \varphi _t \varphi _{tt} \varphi _{ttt}+40 \varphi _t^4 \varphi
	_{tt}+\frac{20}{3} \varphi _{4 t} \varphi _t^2+\frac{\varphi _{6 t}}{3}+\frac{20 }{3} \varphi
	_{tt}^3 =0 ,\\
	&\Leftrightarrow& u_x +\frac{80}{3} u u_t u _{tt} +40 u^4 u_t+\frac{20}{3} u^2 u_{3 t} +\frac{u _{5 t}}{3}+\frac{20 }{3} u_t^3 =0,  \label{248}
\end{eqnarray}	
 We easily establish that ${\cal H}'^r_3 $  is a conserved quantity 
\begin{eqnarray}
\frac{d H'^r_3}{dt} &=& \int \frac{d {\cal H}'^r_3}{dt}  dx = \int u \left[\frac{80}{3} u u_t u _{tt} +40 u^4 u_t+\frac{20}{3} u^2 u_{3 t} +\frac{u _{5 t}}{3}+\frac{20 }{3} u_t^3 \right]dx \\
&=& - \frac{1}{2}\int (u^2)_x  dx =0 ,
\end{eqnarray}
up to the validity of the equation of motion (\ref{248}) and the vanishing of surface terms.

\subsubsection{Integrability from Painlev\'e test }
Next we carry out a Painlev\'e test for the PDE (\ref{248}). The leading order contributions from each of the terms are in this case $u_x \sim w^{\alpha-1}$,   $ u u_{ t }  u_{ tt } \sim  w^{2\alpha-1}$, $ u^4 u_{ t } \sim  w^{5\alpha-1}$, $ u^2 u_{ ttt } \sim  w^{\alpha-3}$, $ u_{5 t} \sim w^{\alpha-5}$,  $ u_t^3 \sim w^{3 \alpha-3}$.  Matching the powers of the third and fifth or sixth term, $ 5 \alpha -1 = \alpha -5 $, we find that $\alpha = -1$. Using this value in the expansion (\ref{Painexpan}) we read off the coefficients in the powers of $w$ and set them to zero
\begin{equation}
	w^{-6}: \quad -20 \lambda _0 w_t \left(\lambda _0^2+2 w_t^2\right) \left(2 \lambda _0^2+w_t^2\right)  =0 .  \label{283a}
\end{equation}	
For the solution $\lambda_0 =i w_t/ \sqrt{2}$  to (\ref{283a}) we find at the next orders
\begin{eqnarray}
	w^{-5}: &&  \,\,   \frac{20}{3} w_t^4 \left(4 \lambda _1 w_t+i \sqrt{2} w_{\text{tt}}\right)=0 \quad \Rightarrow  \quad   \lambda _1 = -\frac{i w_{\text{tt}}}{2 \sqrt{2} w_t},    \label{261xx}   	 \\
	w^{-4}:  &&  \frac{20 i w_t^2}{3}    \left[  4 i \left( 4 w_{tt} +w_t \partial_t  \right) \left( \lambda_1 + \frac{i w_{tt}}{2 \sqrt{2} w_t}   \right)   + 6 \sqrt{2} w_t^2 \left(\lambda _1^2+\frac{w_{tt}^2}{8 w_t^2}\right)    \right]  \Rightarrow \lambda _2  \,
	\text{arbitrary}  ,    \notag   \\
		w^{-3}:  && \,\, \Rightarrow \lambda _3  \,
		\text{arbitrary} ,   \notag \\
	w^{-2}:  && \,\,    \lambda_4 =  \frac{1}{320 w_t^7}   \left\{  80 w_t^3 \left[-2 w_t \left(w_{\text{tt}} 
	\left( \lambda_2 \right)_t   +2 \lambda _2
	w_{ttt}\right)-2 w_t^2 \left(     	\left( \lambda_2 \right)_{tt}      +2 \lambda _3
	w_{tt}\right)  \right. \right. , \notag \\
	&& \left. \left.
	\quad -4 w_t^3     	\left( \lambda_3 \right)_t      +5 \lambda _2 w_{tt}^2    \right]   \right\} -i \sqrt{2} \left\{ 4 w_t \left[ w_t \left(10 w_{4 t} w_{tt}+200 \lambda _2^2 w_t^4-2 w_t
	\left(w_{5 t}+3 w_x\right) \right. \right. \right. \notag  \\
	&& \quad \left. \left. \left.
	+5 w_{ttt}^2\right)  -25 w_{\text{tt}}^2 w_{ttt}\right]+45
	w_{\text{tt}}^4\right\}  ,
	  \notag    \\
	w^{-1}:  &&\,\,     \lambda _5  \,
	\text{arbitrary}  ,      \notag   \\
	w^{0}:  && \,\,      \lambda _6  \,
	\text{arbitrary}   .   \notag       
\end{eqnarray}
This in this case we found the four free parameters $\lambda_2$, $\lambda_3$, $\lambda_5$, and $\lambda_6$, so that together with the fundamental singularity at $\lambda_{-1}$, we match the order of the differential equation and deduce that the system is integrable. Once again we can compare with the necessary condition. The Ansatz (\ref{nescond}) with solution $\lambda_0 = \pm i w_t/ \sqrt{2} $ to the first constraint yields 
\begin{equation}
	\frac{1}{3} \theta  (r-6) (r-5) (r-3) (r-2) (r+1) w_t^5 =0 ,
\end{equation}
which identifies the same powers $r=2,3,5,6$ that we found in our explicit construction.  For the other two solutions $\lambda_0=0$ and $\lambda_0 = \pm i \sqrt{2} w_t  $, we find
\begin{eqnarray}
		\frac{1}{3} \theta  (r-5) (r-4) (r-3) (r-2) (r-1) w_t^5 &=&0,\\
 \frac{1}{3} \theta  (r-8) (r-6) (r-5) (r+1) (r+3) w_t^5 &=&0,
\end{eqnarray}
respectively. In all cases we find five free parameters that may be verified by an explicit construction. 

\section{Exact benign and malevolent solutions and their classical energies}
Next we construct solutions for the modified rotated KdV equations (\ref{rotKdV}) and those resulting from interpreting the higher charges of the rotated system as Hamiltonians (\ref{24811}) and  (\ref{248}). With regard to the question of whether the characteristic features of HTDT show up at the classical level, we are especially interested in periodic benign solutions in time with sufficiently many free parameters to satisfy the initial conditions of the rotated Cauchy problem. For some examples we will also constrain the real $x$-axis and thus include boundaries, so that the Cauchy problem is extended to an initial-boundary value problem. Assuming travelling wave solutions these equations may be integrated out and with appropriate assumptions periodic solutions can be obtained. Here we use instead, except for the linear $(n=2)$-case, the extended Jacobian elliptic function expansion method \cite{liang03ex} that yields exact periodic solutions in a straightforward manner. Important for our purposes here is that this approach also exhibits the different types of dispersion relations. 

\subsection{Exact solutions for the rotated $n=2$-mKdV equations of motion} 
In constructing solutions to the equations of motion (\ref{rotKdV}) we start with the $(n=2)$-case, which is straightforward to solve as it is linear and can be solved explicitly by elementary methods, but is still instructive in exhibiting some key characteristics. Factorising $u(x,t)=f(x)g(t)$, equation  (\ref{rotKdV}) is converted into the two equations
\begin{equation}
	\frac{f_x}{f}=- \lambda, \qquad  \frac{2 g_t}{g} + \frac{g_{ttt}}{g} = - \lambda ,
\end{equation}
with $\lambda \in \mathbb{C}$ being a constant. Solving both equations, the general solution then simply becomes
\begin{equation}
	u_2(x,t) =  \left( c_1 e^{- m_1 t} + c_2 e^{- m_2 t} +c_3 e^{- m_3 t} \right) c_4 e^{- \lambda x}, \label{soln2}
\end{equation}
with $m_i$ denoting the three roots of $t^3 + 2 t + \lambda =0$ and $c_i$ the required integration constants. A similar solution is obtained when assuming a travelling wave solution of the form $u(x,t) = u(\zeta)$ where $\zeta := x - c t$. We can then integrate out the equation directly and subsequently once more when multiplied by $u_\zeta$ obtaining
\begin{equation}
      \frac{du}{d \zeta} =   \sqrt{ \frac{2}{c^3}   \left[  \left(  \frac{1}{2} - c      \right) u^2  - c_1 u - c_2   \right]}, \label{trav2}
\end{equation}
with integration constants $c_1 , c_2$. Equation (\ref{trav2}) is easily integrated out producing a similar solution as reported in (\ref{soln2}). 

\begin{figure}[h]
	\centering         
	\begin{minipage}[b]{0.31\textwidth}      
		\includegraphics[width=\textwidth]{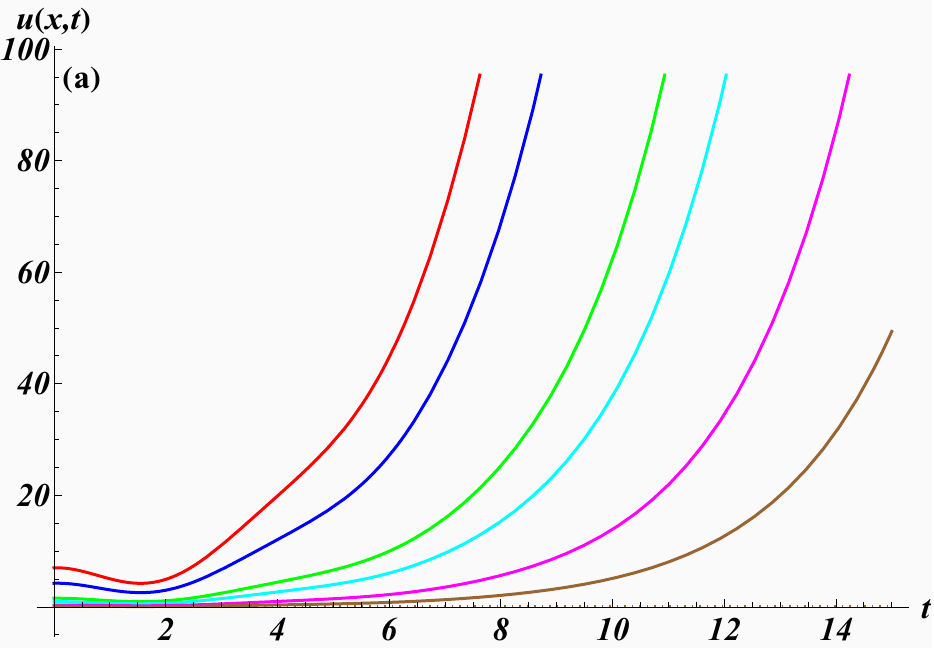}
	\end{minipage}   
	\begin{minipage}[b]{0.31\textwidth}      
		\includegraphics[width=\textwidth]{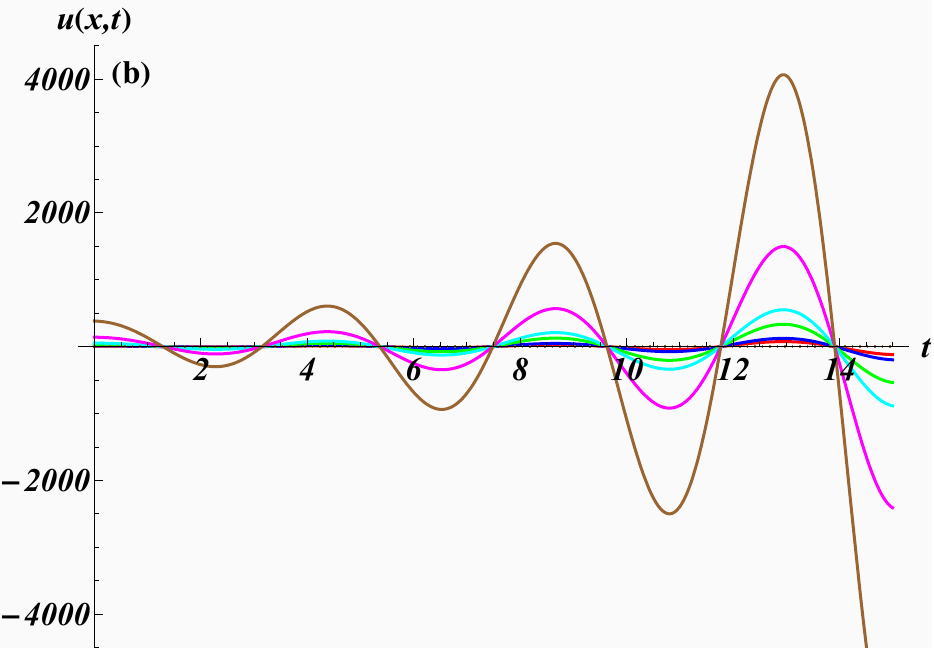}
	\end{minipage} 
		\begin{minipage}[b]{0.36\textwidth}      
		\includegraphics[width=\textwidth]{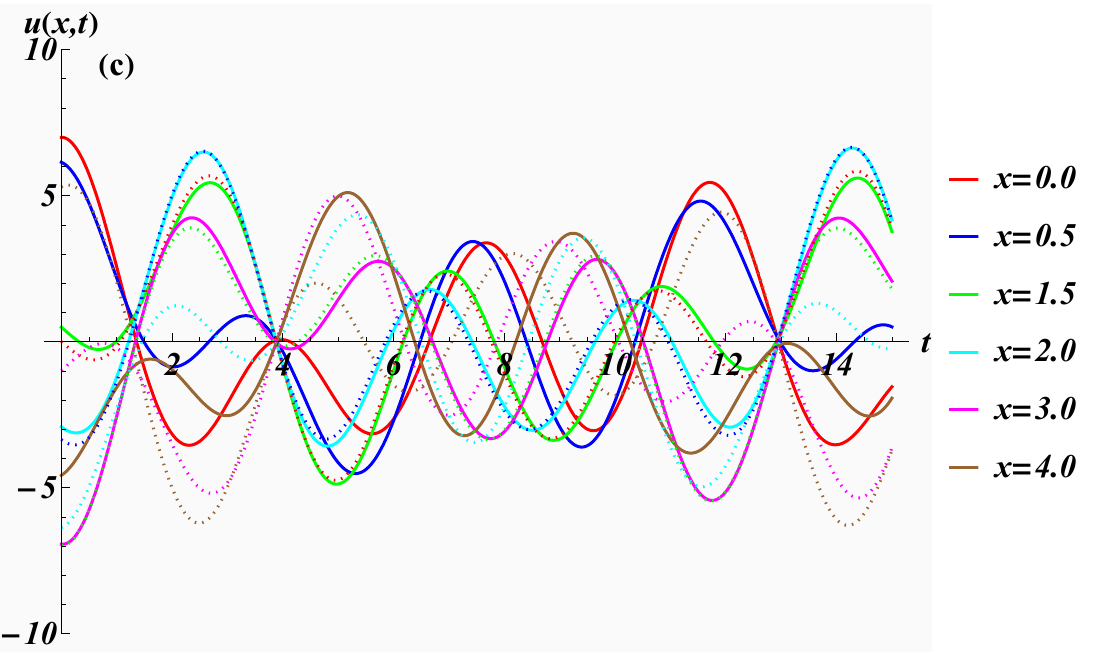}
	\end{minipage} 
	\caption{Classical solutions for the $n=2$ rotated modified KdV equation at different locations $x$ as functions of time $t$ with $c_1=3$, $c_2=c_3=2$, $c_4=1$, $m_1 \in \mathbb{R}$, $m_2 =m^*_3 \in \mathbb{C}$ for $\lambda=1$, $\lambda=-1$, $\lambda=i$ in panels (a), (b) and (c), respectively. In panel (c) real parts are depicted as solid lines and imaginary parts as dotted lines.} 
	\label{figsoln2}
\end{figure}

 Depending on the values of $\lambda$ we encounter malevolent exponentially and oscillatory divergent solutions as well as benign complex solutions that stay finite when time evolves, see figure \ref{figsoln2}. Dissipative solutions of similar type were previously found for extended versions of the Pais-Uhlenbeck oscillators on the classical and quantum level \cite{nest07inst,szegleti20diss,sandersdiss}. In section 5 we will comment on the field theoretical version. 
 
Notice that even having a benign solution and despite having as many constants in the solution (\ref{soln2}) as the order of the differential equation in $t$, we can not construct from them an exact solution to the rotated Cauchy problem, as the initial profile functions $u_2(x,0)$, 	$(u_2)_t(x,0)$ and $(u_2)_{tt}(x,0)$ can  not be chosen arbitrarily, but are linearly dependent of the form $\sim e^{- \lambda x}$.

The energies of the solutions are computed from the densities in (\ref{Haminu}).  For real values of $\lambda$ the energies become infinite when integrating over the entire real axis and when $\lambda \in i \mathbb{R}$ we obtain vanishing energies on finite intervals $E=\int_{- \pi/2}^{\pi/2} {\cal H}'^r(u_2) dx =0$.

\subsection{Exact solutions for the rotated $n=3$-mKdV equations of motion} 

Next we consider the standard KdV equation (\ref{stanKdV}) with $n=3$ in order to exhibit the key differences between the original and the rotated version. Making an Ansatz for the solution of the original version in (\ref{stanKdV}) in terms of the Jacobi elliptic function $\text{sn}(z |m)$ with parameter  $m \in [0,1]$, and at this point unknown constants $\alpha$, $k$, $\omega$, 
\begin{equation}
	u_3(x,t) =\alpha   \,\text{sn}^2(k x+t \omega |m),  \label{AnKdV}
\end{equation}
we obtain, upon substitution into original KdV equation (\ref{stanKdV}), the constraining equation
\begin{eqnarray}
	&&2 \alpha  \text{cn}(k x+t \omega |m) \text{dn}(k x+t \omega |m) \text{sn}(k x+t \omega |m) \\
    && \qquad \qquad \times 	\left[ \omega- 4 k^3 (1+m)  + 6 k \left(\alpha +2 k^2 m\right) \text{sn}^2 (k x+t \omega |m)
	\right] =0 .\notag
\end{eqnarray}
The last factor vanishes when we set $\alpha =-2 m k^2 $ and the dispersion relation to $\omega(k)=4 k^3 (1+ m)$. We will make use of the real and complex periodicities
\begin{equation}
	u_3\left(x,t\right) = u_3\left(x +\frac{2}{k} K,t\right)= u_3\left(x +\frac{2}{k} i K',t\right)
	=  u_3\left(x ,t+\frac{2}{\omega} K\right) = u_3\left(x ,t+\frac{2}{\omega}i K'\right),
\end{equation}
with $K:=K(m)$, $K':=K(1-m)$ denoting the complete elliptic integrals of the first kind. In the limit $m \rightarrow 1$ this solution reduces to the well-known solution $u_3(x,t) =-2 k^2 \text{tanh}^2\left(8 k^3 t+k x\right)$. Similar solutions may be obtained from different starting points akin to (\ref{AnKdV}) involving other Jacobi elliptic functions. 

For the rotated KdV equation the Ansatz (\ref{AnKdV}), when substituted into (\ref{rotKdV}), yields the constraint
\begin{eqnarray}
	&&2 \alpha  \text{cn}(k x+t \omega |m) \text{dn}(k x+t \omega |m) \text{sn}(k x+t \omega |m)  \\
	&& \qquad \qquad \times	\left[k-4 (1+m) \omega ^3+ 6 \omega  \left(\alpha +2 m \omega ^2\right) \text{sn}^2(k x+t \omega |m)\right], \notag
\end{eqnarray}	
so that when solving the dispersion relation for $\omega$, we obtain three distinct solutions
\begin{equation}
	u_{3r}^{(n)}(x,t) =\alpha^{(n)}   \,\text{sn}^2 \left[  k x+t\omega^{(n)}   |m\right] , \quad  \alpha^{(n)} := -2 m \left[\omega^{(n)} \right]^2, \quad \omega^{(n)} :=   \left[ \frac{k}{ 4  + 4 m}  \right]^{1/3}   \tau^{n} \
	\label{rotsolKdV}
\end{equation}
for $n=1,2,3$ with $\tau := e^{2 \pi i/3}$ denoting the third primitive root of unity.

\begin{figure}[h]
	\centering         
	\begin{minipage}[b]{0.52\textwidth}      
		\includegraphics[width=\textwidth]{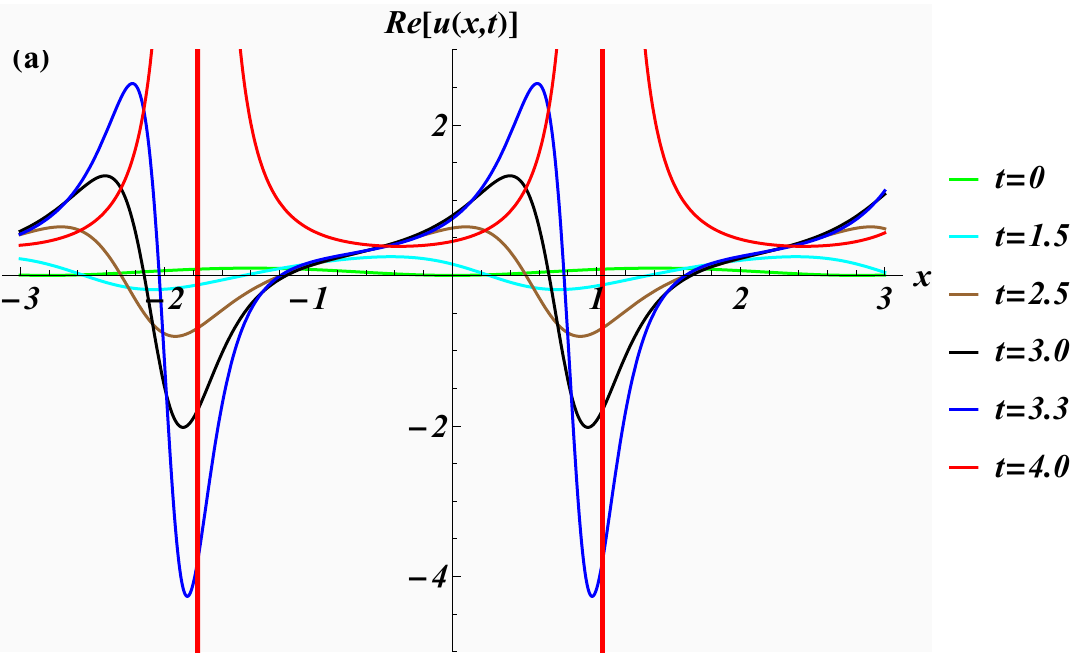}
	\end{minipage}   
	\begin{minipage}[b]{0.45\textwidth}      
		\includegraphics[width=\textwidth]{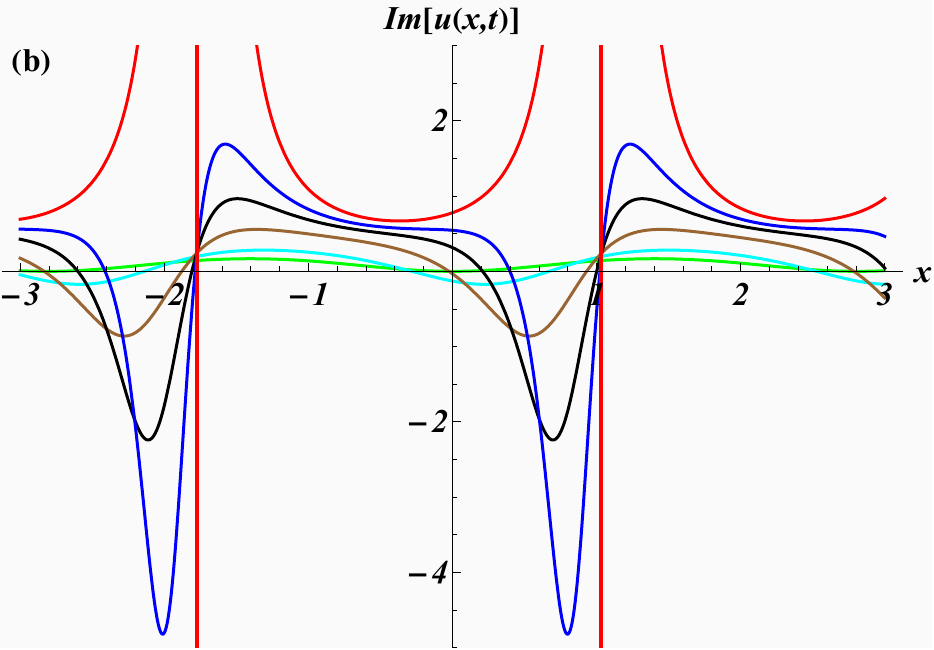}
	\end{minipage} 
	\caption{Classical malevolent solutions $u_{3r}^{(1)}(x,t)$ at different times for the $n=3$ rotated modified KdV equation with $k=1.2$, $m=0.25$ and singularities at the points $	(\hat{x}_{-1,0}^{(1)}, t_{-1,0}^{(1)}) = (1.0376, 4.0070)$,  $	(\hat{x}_{-1,-1}^{(1)}, t_{-1,0}^{(1)}) = (-1.7720, 4.0070)$.   } 
	\label{malevolentFig}
\end{figure}

The complex solutions $u_{3r}^{(1)}(k>0) = u_{3r}^{(2)}(k<0)= [u_{3r}^{(2)}(k>0) ]^* = [u_{3r}^{(3)}(k<0) ]^*$ are malevolent as they have singularities at $(k x+t\omega^{(1,2)} ) = \ell i K'$ for $\ell \in \mathbb{Z}$, i.e. at the points 
\begin{eqnarray}
	\hat{x}_{\ell,\mu}^{(n)} &=& -\frac{\ell}{k} \cot \left(\frac{2 \pi  n}{3}\right)  K' + \frac{2 \mu K}{k}, \qquad  \qquad \qquad  \qquad \qquad  \,\, n=1,2 \,\, \mu,\nu 
	\in \mathbb{Z} ,  \\
	\hat{t}_{\ell,\nu}^{(n)} &=&  \frac{ 2^{2/3} }{k^{1/3}}   (m+1)^{1/3}  K \csc \left(\frac{2 \pi  n}{3}\right)
	\left[ \ell K' + 2 \nu  \tan\left(\frac{2 \pi  n}{3}\right) K  \right],
	\label{singu}
\end{eqnarray} 
that are reached in finite time $t$. Figure \ref{malevolentFig} shows how a typical malevolent solution runs into a singularity as time evolves. 

In contrast, the real solutions $u_{3r}^{(3)}(k>0) = u_{3r}^{(1)}(k<0)$ are oscillatory and of a benign nature. However, this does of course not yet mean that we can solve the rotated Cauchy problem as for this we need accommodate three arbitary independent function $u$, $u_t$ and $u_{tt}$ as initial profiles at $t=0$. 
In order to obtain the right amount of free functions we have to combine three of the solutions. As our equation is nonlinear we can of course not simply add them, but need to do this by means of a generalised superposition as obtained from a B\"acklund transformation. In appendix A we explain how to achieve that.

\begin{figure}[h]
	\centering         
	\begin{minipage}[b]{0.49\textwidth}      
		\includegraphics[width=\textwidth]{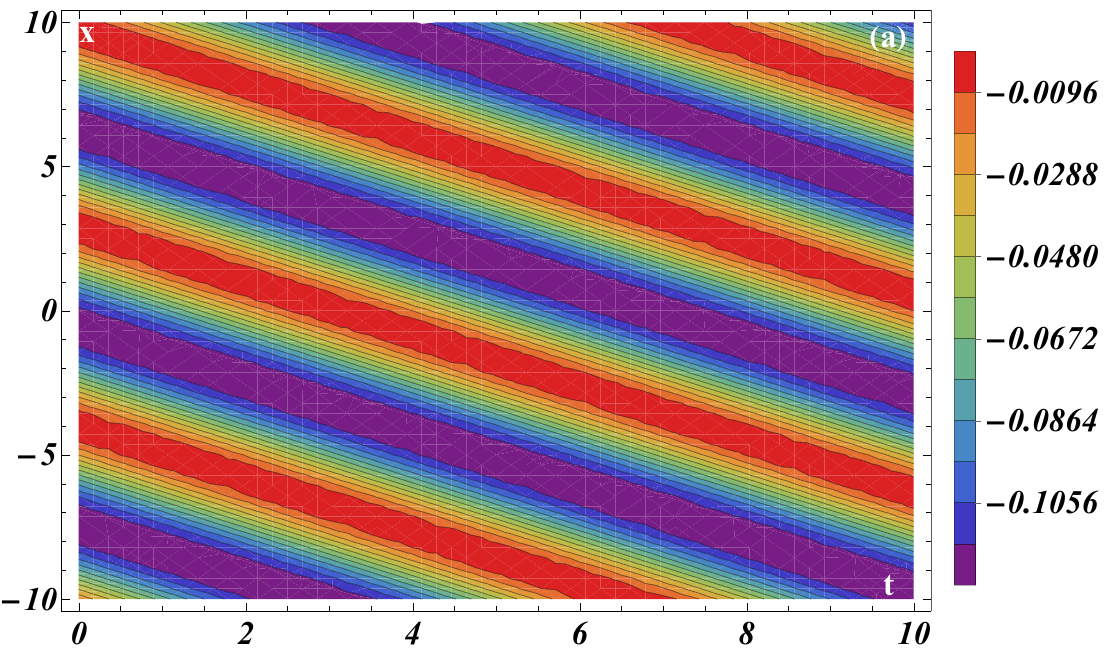}
	\end{minipage}   
	\begin{minipage}[b]{0.49\textwidth}      
		\includegraphics[width=\textwidth]{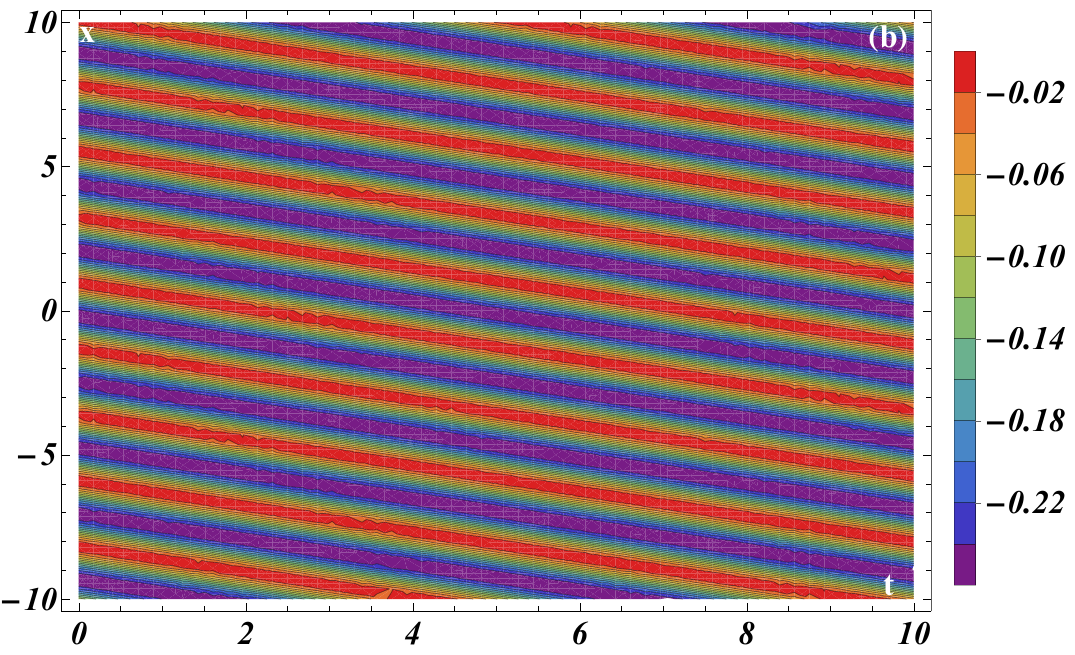}
	\end{minipage} 
	\caption{Classical superimposed solutions (\ref{superposu}) for the $n=3$ rotated modified KdV equation with $\omega = \omega^{(3)} $, $\alpha_{1}= 1/2$, $\alpha_{2}= 3/2$, $m=0.3$, $\lambda = 0$ for $k=1/2$ and $k=3/2$ in panels (a) and (b), respectively.} 
	\label{superab}
\end{figure}

 As seen in figure \ref{superab} when taking $\omega = \omega^{(3)}$ the solutions are benign. Using the three parameters $\lambda$, $\alpha_1$, $\alpha_2$, we can adjust these solutions to generate many initial value profiles for the rotated Cauchy problem thus creating subset in the functional parameter space with exact solutions of benign nature.  In \cite{Smilga6} it was argued that for the KdV-system some of the exact solutions would blow up as they evolve in time and doubt was raised about the existence of benign solutions. Here we have explicitly shown that benign solutions can easily be constructed and combined in such a way that the Cauchy problem is solved for some specific initial profile functions. It remains unclear whether the Cauchy problem can be solved in complete generality.

Next we compute the energies of these solutions in one period $x \in [a - K/k, a+ K/k]$  for arbitrary $a$ associated to the Hamiltonians $H_0$ and $H'^r$. Setting $a=0$, we find real energies for the solution $u_3$ of the original KdV equation
\begin{eqnarray}
	E_3(k,m) &=& \int_{-K/k}^{K/k} {\cal H}_0 (u_3) dx,   \\
	&=& \frac{16}{5} k^5 \left[    \left( m^2+2 m+2\right) K-\left( 2 m^2+3 m+2  \right) K''  \right],  \notag
\end{eqnarray}
with $K''=E(m)$ denoting the complete elliptic integrals of the second kind. As expected, since $H_0$ is conserved in time the energies for particular solutions are indeed time-independent. 

For the solutions $u_{3r}^{(n)}$ of the rotated KdV solution we obtain
\begin{equation}
	E_{3r}^{(n)}(k,m) = \int_{-K/k}^{K/k}    {\cal H}'^r \left[  u_{3r}^{(n)}   \right] dx
	=   \frac{1}{3} k \left[2 K'' - \frac{2+m }{1+m} K \right].
\end{equation}
Notice that the energies are identical, real for all values of $n$ and finite, except in the limit $m \rightarrow 1$, for which we have
\begin{equation}
	\lim_{m \rightarrow 1} E_3(k>0,m), -E_3(k<0,m) , -E_{3r}^{(n)}(k>0,m) , E_{3r}^{(n)}(k<0,m) \rightarrow \infty .
\end{equation}
 The reality of the energy is easily understood by noting that $ {\cal H}'^r \left[  u_{3r}^{(n)}(x,t=0) \right] \in  \mathbb{R}$ and the fact that the energy is a conserved quantity in time. Alternatively, by noting that for two different solutions $u_1$ and $u_2$ the modified ${\cal CPT}$-symmetry: $u_1(x,t) \rightarrow u_2^*(-x,-t) $,  $H'^r(u_1) = [H'^r(u_2)]^*$ is realised pairwise for $u_{3r}^{(n)}$ for the same energy solutions, we can utilise the arguments from \cite{fring2020BPS} to ensure the reality of the energy.

\begin{figure}[h]
 	\centering         
 	\begin{minipage}[b]{0.49\textwidth}      
 		\includegraphics[width=\textwidth]{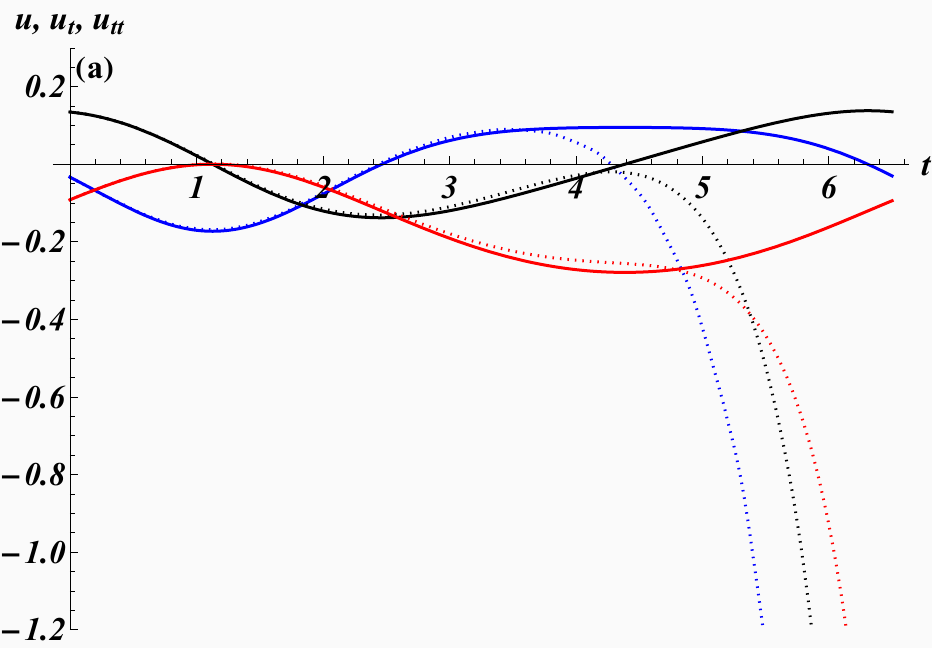}
 	\end{minipage}   
 	\begin{minipage}[b]{0.49\textwidth}      
 		\includegraphics[width=\textwidth]{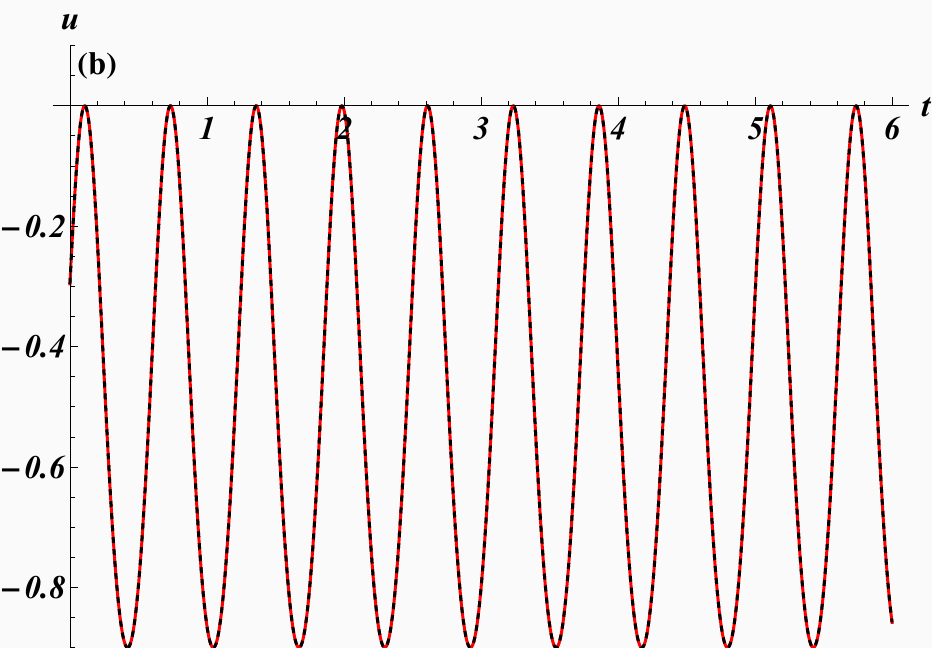}
 	\end{minipage} 
 	\caption{ Exact (solid lines) versus numerical solution (dotted lines) of the rotated and unrotated KdV Cauchy problem in panel (a) and (b), respectively. In panel (a) $u$, $u_t$, $u_{tt}$ are depicted in red, black and blue, respectively. In panel (b) the exact solution is depicted in red and the numerical solution in black. The sample values are $k=1$, $m=0.45$ and $x=3$ in both panels.}
 	\label{exnumKdV}
 \end{figure}

 We finish this section with a discussion of the stability of these exact benign solutions. In principle, a rigorous analytical analysis up to the standard of the unrotated Cauchy problem would be highly desirable, but a good insight can already be obtained from a linearisation \cite{Smilga6} and a numerical analysis as was argued in \cite{Smilgaacta}. Since in any numerical analysis one encounters inaccuracies, simply due to the limited precision of any numerical method, one may compare the exact solution with its numerical solution and interpret the latter as a perturbed version of the former. In figure \ref{exnumKdV} panel (a) we plot the real exact solution $u_{3r}^{(3)}$ versus the numerical solution of the rotated Cauchy problem obtained by using as the initial profile functions the exact solutions, i.e. $u(x,0)=u_{3r}^{(3)}(x,0)$, $u_t(x,0)=\left[ u_{3r}^{(3)}(x,0)\right]_t$ and  $u_{tt}(x,0)=\left[ u_{3r}^{(3)}(x,0)\right]_{tt}$. (Strictly speaking we are solving an initial-boundary value problem as technically we do not cover the entire real axis.) We observe that already before even one period in time is finished the numerical solution deviates drastically from the exact solution and rapidly develops a singularity. Thus the exact benign solution is malevolently unstable according to the above characterisation. In contrast,
 comparing the exact solution $u_{3}(x,t)= -2 k^2 m \text{sn}\left[ \left.4 (m+1)  k^3 t +x k\right|m\right]$ of the original unrotated equation with the numerical solution computed with only one initial value function $u(x,0)=u_{3}(x,0)$, we observe that the numerical solution follows precisely the exact solution as seen in panel (b) of the same figure. The agreement holds essentially for arbitrary large time, so that the solution of the original Cauchy problem is stable and remains benign. 
 
\subsection{Exact solutions for the rotated $n=4$-mKdV equations of motion} 

In order to solve the rotated modified KdV equation (\ref{rotKdV}) with $n=4$ we have to slightly alter the Ansatz (\ref{AnKdV}) and reduce the order of the Jacobi elliptic function. Proceeding in the same manner, substituting 
\begin{equation}
	u_{4r}(x,t) =\alpha   \,\text{cn}(k x+t \omega |m) ,  \label{AnmKdV}
\end{equation}
into (\ref{rotKdV}) yields the constraint 
\begin{equation}
	\alpha  \text{dn}(k x+t \omega |m) \text{sn}(k x+t \omega |m) \left[ 6 \omega  \left(m
	\omega ^2-2 \alpha ^2\right) \text{cn}(k x+t \omega |m)^2-k+(1-2 m) \omega ^3\right],
\end{equation}
leading to the solutions
	\begin{equation}
	u_{4r}^{(n)}(x,t) =\alpha^{(n)}   \,\text{cn} \left(k x+t\omega^{(n)}   |m\right) , \quad  \alpha^{(n)} := \frac{\sqrt{m} \omega^{(n)}}{\sqrt{2}}  , \quad \omega^{(n)} :=   \left( \frac{k}{ 1 -2 m}  \right)^{1/3}   \tau^{n}.
\end{equation}
The complex solutions are once again of malevolent nature, whereas the real solutions $u_{4r}^{(1)}$ for $m>1/2$ and $u_{4r}^{(3)}$ for $m<1/2$ are benign. Combining the solutions $u_{4r}$ according to the Miura transformation (\ref{Miura}) produces indeed a new solution to the rotated $n=3$-mKdV equation.

For the standard solution $u_4$ of the original modified KdV equation we compute
\begin{equation}
	E_4(k,m) =  \int_{-K/k}^{K/k} {\cal H}_0 (u_4) dx
	= -\frac{1}{6} k^3 \left\{  [ m (3 m-4) +1] K+(2 m-1) K''  \right\}, 
\end{equation}
which for the $\sech$-solution becomes $\lim_{m \rightarrow 1} E_4(k,m) = -1/6 k^3$. In turn for the solutions $u_{4r}^{(n)}$ of the rotated mKdV equation we obtain
\begin{equation}
	E_{4r}^{(n)}(k,m) = \int_{-K/k}^{K/k} {\cal H}'^r \left[  u_{4r}^{(n)}   \right] dx
	=\frac{  \omega ^{(n)}}{4 m-2}  \left\{  \left[ (4-3 m) m-1\right]K+(1-2 m) K'' \right\} ,
\end{equation}
which are never real for all $n$. We have either $	E_{4r}^{(3)} \in \mathbb{R} $ and $	E_{4r}^{(1)} = [E_{4r}^{(2)}]^*  \notin \mathbb{R} $ when $k>0, m<1/2$; $k<0, m>1/2$ or $	E_{4r}^{(1)} \in \mathbb{R} $ and $	E_{4r}^{(2)} = [E_{4r}^{(3)}]^*  \notin \mathbb{R} $ when $k>0, m>1/2$; $k<0, m<1/2$. In the trigonometric limit we obtain $\lim_{m \rightarrow 1} E_{4r}^{(n)}(k,m) = -\tau ^n (-k)^{1/3}$. In the case the ${\cal CPT}$-symmetry is broken for the complex solutions $u_{4r}^{(n)}$  and leads to complex conjugate energy solutions.

\begin{figure}[h]
	\centering         
	\begin{minipage}[b]{0.50\textwidth}      
		\includegraphics[width=\textwidth]{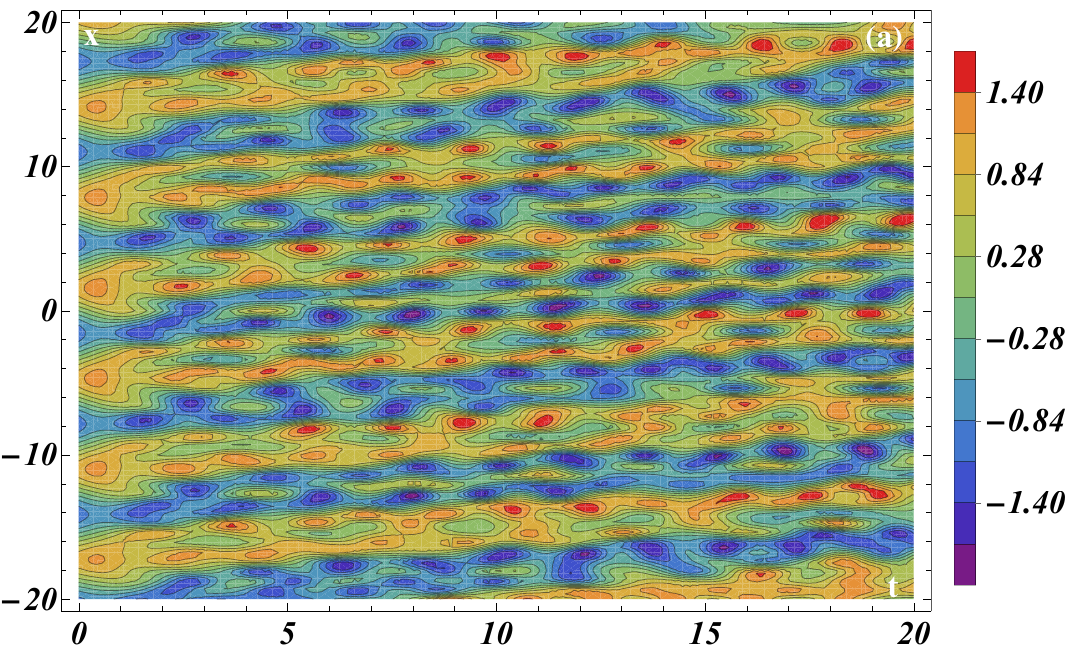}
	\end{minipage}   
	
	\begin{minipage}[b]{0.49\textwidth}      
		\includegraphics[width=\textwidth]{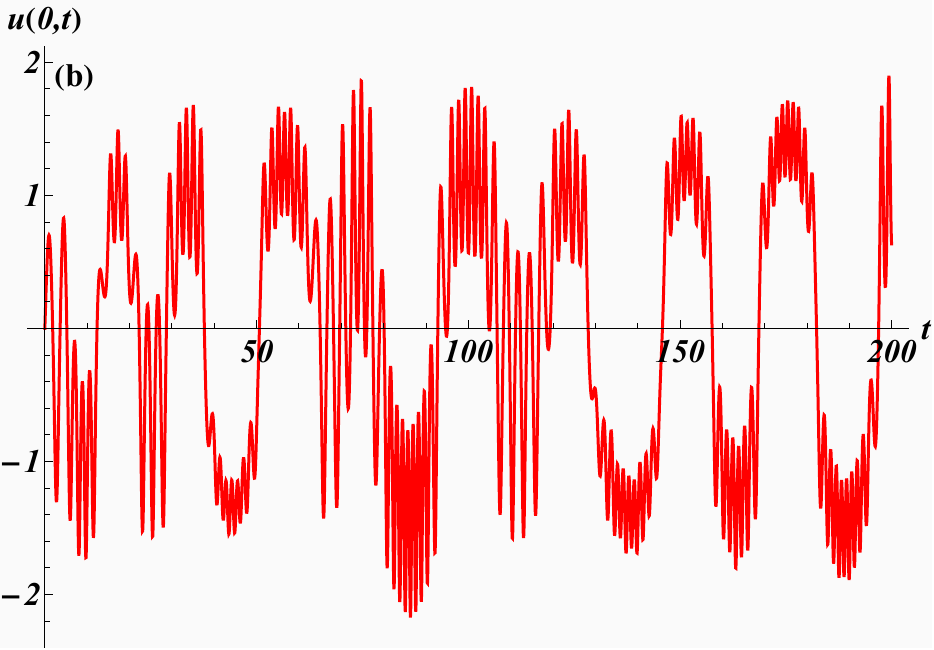}
	\end{minipage} 
	\begin{minipage}[b]{0.49\textwidth}      
		\includegraphics[width=\textwidth]{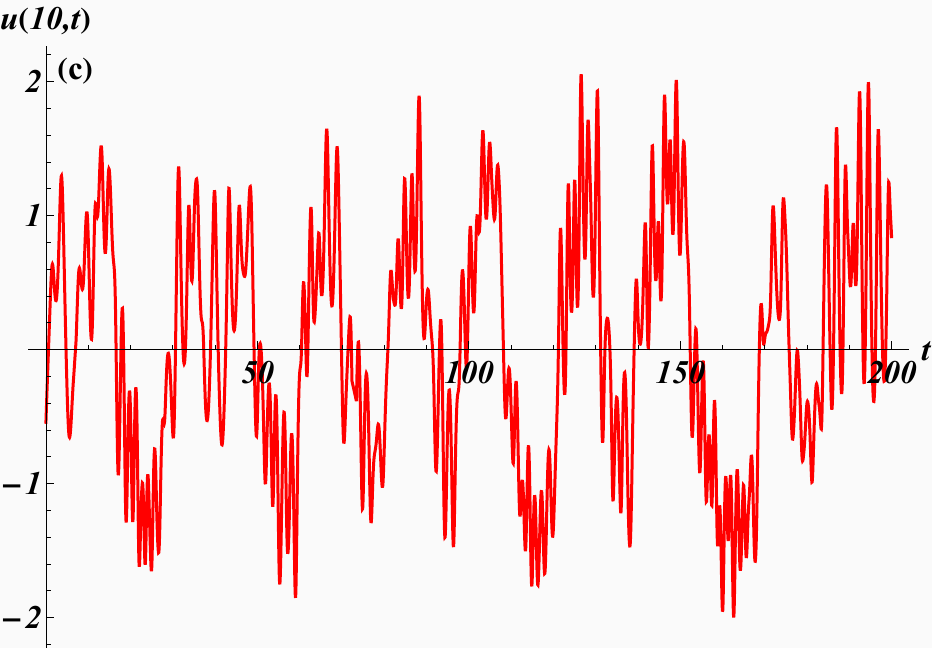}
	\end{minipage} 
	\caption{Classical solutions to the initial-boundary value problem for the $n=4$ rotated modified KdV equation with boundary value functions (\ref{bvfunc}) in panels (a) and selected timelines  for fixed values of $x$} in panels (b) and (c).
	\label{BIVBmKdV}
\end{figure}

To address the rotated Cauchy problem and find more possibilities for initial profile functions we could in principle carry out a similar analysis as in the previous subsection by constructing exact nonlinearly superpositioned solutions involving several parameters from a rotated mKdV B\"acklund transformation following \cite{wadati74B}. However, in that case the analogue equations to (\ref{A4}) and (\ref{A5}) can not be solved analytically when the seed solution is taken to be a Jacobi elliptic function. A similar problem occurs when trying to build such a solution form an alternative version of the mKdV B\"acklund transformation when following \cite{schiff94B}. Alternatively, one could combine several solutions using Darboux–Crum transformations \cite{arancibia14sol}, but instead we will consider here a numerical study of the corresponding initial-boundary value problem. Thus, we consider (\ref{rotKdV}) on the domain $t \in [0,\infty)$, $x \in [-X,X]$ with 
\begin{equation}
     u(x,0)=f_1(x), \,\,\,  u_t(x,0)=f_2(x), \,\,\,  u_{tt}(x,0)=f_3(x), \,\,\, u(\pm X,t) = a_\pm(t),
\end{equation}
where the functions $f_i(x)$ are arbitrary. Consistency of the constraints requires that $a_\pm(0)=f_1(\pm X)$,  $(a_\pm)_t(0)=f_2(\pm X)$ and  $(a_\pm)_{tt}(0)=f_3(\pm X)$. As a compatible set of solutions for the initial-boundary value functions we take here for instance
\begin{equation}
f_1(x)=\sin(x), \,\,\, f_2(x)=\cos(2 \pi x/X), \,\,\, f_3(x)=0, \,\,\, a_\pm(t)= \sin(t) \pm \sin(X).
\label{bvfunc}
\end{equation}
In figure \ref{BIVBmKdV} we depict the numerical solutions for $u(x,t)$ for a particular interval together with some typical timelines.

We observe that solutions appear to remain benign and do not develop any singularities as time evolves. This confirms the findings in \cite{Smilga6} for a different set of boundary value functions. 

\begin{figure}[h]
	\centering         
	\begin{minipage}[b]{0.49\textwidth}      
		\includegraphics[width=\textwidth]{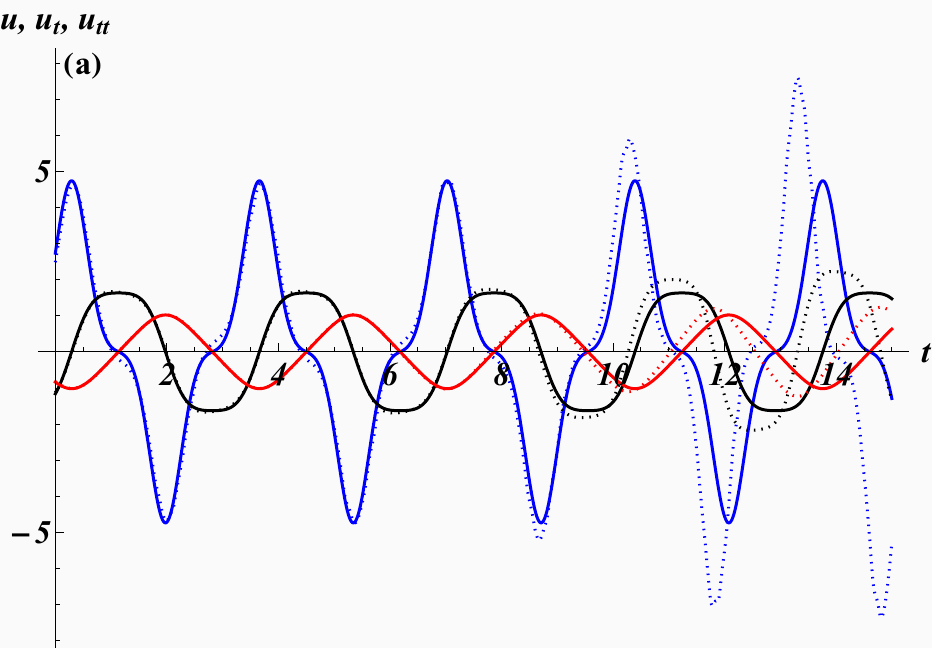}
	\end{minipage}   
	\begin{minipage}[b]{0.49\textwidth}      
		\includegraphics[width=\textwidth]{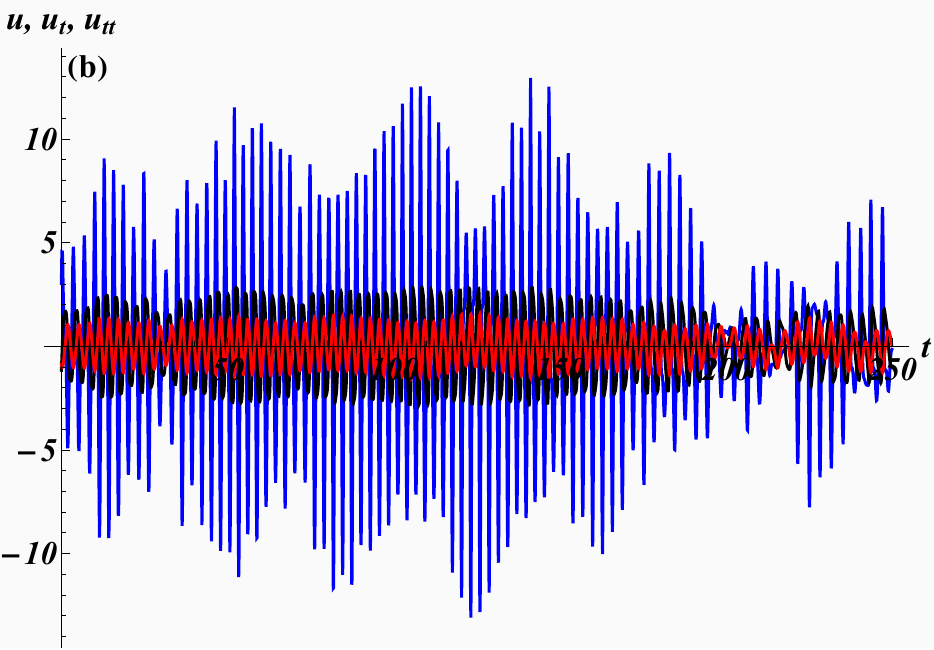}
	\end{minipage} 
	\caption{ Exact (solid lines) versus numerical solution (dotted lines) of the rotated mKdV Cauchy problem for small times in panel (a). Numerical solution of the rotated mKdV Cauchy problem for large times in panel (b). The functions $u$, $u_t$, $u_{tt}$ are depicted in red, black and blue, respectively. The sample values are $k=1$, $m=0.45$ and $x=3$ in both panels.}
	\label{exnummKdV}
\end{figure}

The stability of these solutions is examined closer in the same way as in the previous subsection,  see figure \ref{exnummKdV}.  In this case we may directly compare with a similar calculation carried out in \cite{Smilgaacta}. First of all we notice that the ``high frequency noise" observed in figures 4 and 5 in there is absent in our calculation even for the higher derivatives. This is simply due to a different numerical method used. (Here we used a higher-order Runge Kutta method.) In panel (a) we observe that also in this case the solution of the rotated Cauchy problem is unstable, as the numerical solution deviates from the exact solution after about two and a half periods, depending on the values chosen. However, unlike as in the case of the KdV equation we observe here that the solution does not diverge and stays finite even for large times as seen in panel (b). Thus this solution is unstably benign, which confirms the findings made in  \cite{Smilgaacta}, see figure 6 therein. Once again, we find that the numerical solution for the unrotated case follows closely the exact solution similarly as found for the KdV equation depicted in panel (b) of figure \ref{exnumKdV}.

\subsection{Exact solutions for the rotated higher charge $n=3,4$-mKdV systems} 

Periodic solutions to the $n=3$ equation of motion (\ref{24811}) obtained from the higher charge Hamiltonian (\ref{336z}) can be obtained in the same way resulting to
\begin{equation}
	\hat{u}_{3r}^{(n)}(x,t) =\alpha^{(n)}   \,\text{cn}^2 \left(k x+t\omega^{(n)}   |m\right) , \,\, \alpha^{(n)}:=  2m  \omega^{(n)} , \,\, \omega^{(n)} :=
	\left(\frac{3 k}  { 8(7 m-7 m^2-2)}\right)^{1/5}  \hat{\tau}^{n} ,
\end{equation}
for $n=1,\ldots,5$ with $\hat{\tau} := e^{2 \pi i/5}$ denoting the fifth primitive root of unity. As above, the complex solutions diverge at finite time, whereas the real solutions  $\hat{u}_{3r}^{(5)}(x,t), k< 0 = \hat{u}_{3r}^{(2)}(x,t), k> 0$ are benign.

The energies resulting from the higher charge Hamiltonian (\ref{336z}) for these solutions are computed to
\begin{eqnarray}
	\hat{E}_{3r}(k,m) &=&  \int_{-K/k}^{K/k} {\cal H}_3^r \left[ \hat{u}_{3r}^{(n)} \right]  dx \\
  	&& \!\!\!\! \!\!\!\! \!\!\!\!  \!\!\!\! \!\!\!\! \!\!\!\!  = \frac{16  \left[\omega^{(n)}\right]^8}{9 k} \left[   (m-1) \left(45 m^3-73 m^2+40 m-8\right) K+4 (2 m-1) \left(7 m^2-7 m+2\right) K''  \right],  \notag
\end{eqnarray}
which is real only for the real solutions.

Solutions to (\ref{248}) are obtained in the same way resulting to
	\begin{equation}
	\hat{u}_{4r}^{(n)}(x,t) =\alpha^{(n)}   \,\text{cn} \left(k x+t\omega^{(n)}   |m\right) , \,\, \alpha^{(n)}:= \sqrt{\frac{m}{2}} \omega^{(n)} , \,\, \omega^{(n)} :=
	\left(\frac{3 k}{6 m-6 m^2-1}\right)^{1/5}  \hat{\tau}^{n} , \label{omegan}
\end{equation}
for $n=1,\ldots,5$, with $\hat{u}_{4r}^{(5)}(x,t), k>0 = -\hat{u}_{4r}^{(2)}(x,t), k<0 $ being real benign solutions and the remaining ones complex divergent in time. 

The energies resulting from the higher charge Hamiltonian (\ref{371z}) for these solutions are 
\begin{eqnarray}
	\hat{E}_{3r}(k,m) &=&  \int_{-K/k}^{K/k} {\cal H}_3^r \left[ \hat{u}_{4r}^{(n)} \right]  dx \\
	&=&  \frac{  \left[\omega^{(n)}\right]^6}{126 k }
	  \left\{  \left[ 7 \left(14 m^2-14 m-1\right) -4 (2 m-1) \left(6 m^2-6 m-19\right) \left[\omega^{(n)}\right]^2 \right] K''   \right.  \notag  \\
	 && 
	  + \left. (m-1)   \left[ 7 \left(30 m^2-22 m-1\right)  + 2  \left(12 m^2+45 m-38\right) \left[\omega^{(n)}\right] ^2         \right]       K  \right\} ,  \notag
\end{eqnarray}
which is also real only for the real solutions. 

So, in the $n=3$-theory the energies, resulting from the Hamiltonians are always real, even for the complex solutions, but for the $n=3$ higher charges and all charges in the $n=4$-theory, only the real solutions posses real energies. 

\section{Quantization}

We follow here the approach elaborated on in \cite{weldon03quant} for higher time-derivative generalizations of the Klein-Gordon theory. The initial step of the scheme consists of Fourier transforming the scalar fields 
\begin{equation}
	\varphi(x,t) = \frac{1}{2 \pi} \int dk \phi(t,k) e^{i k x} ,
\end{equation}   
and subsequently quantizing the theory involving the fields $ \phi(t,k)$ for fixed wave numbers $k$. For the scheme to be applicable the resulting $k$-dependent Lagrangian is required to be of the general form
\begin{equation}
	L_k = \frac{1}{2} \sum_{n=0}^N C_n(k)   \left( \frac{d^{n} \phi(t,k)}{dt^{n}} \right)^2 .  \label{Weldonform}
\end{equation}  
In general, the higher time-derivative Lagrangians considered above contain products of three or more of the fields and their derivatives, such that they are not of the form (\ref{Weldonform}). However, the Fourier transformed Lagrangian of the rotated version (\ref{L235}) for $n=2$ reads 
\begin{equation}
	L^r_k =\frac{1 }{2}  \left( i k \phi_t \phi + 2 \phi_t^2 - \phi_{tt}^2  \right)= \frac{1}{2} \sum_{i,j=0}^2 \tilde{C}_{ij} \frac{d^{i} \phi}{dt^{i}}  \frac{d^{j} \phi}{dt^{j}} = \frac{1}{2} \Phi^\intercal \tilde{C} \Phi,  \label{Weldon}
\end{equation}  
with 
\begin{equation}
\tilde{C}:= \begin{pmatrix}
	0 & \frac{i k}{2} & 0 \\
\frac{i k}{2}   & 2   & 0 \\
0 & 0 & -1 
\end{pmatrix}, 
\qquad 
\Phi:= \begin{pmatrix}
\phi \\
	\phi_t \\
	\phi_{tt}
\end{pmatrix}, 
\end{equation}  
which can be transformed to the variant (\ref{Weldonform}) by means of an orthogonal transformation $S$. The transformed Lagrangian in the fields $\psi$ is derived as
\begin{equation}
	L^r_k =\frac{1 }{2}  \left\{ -\psi^2 + \left[ 1-\sqrt{1-\frac{k^2}{4} } \right] \psi_t^2 +\left[1+\sqrt{1-\frac{k^2}{4} } \right]  \psi_{tt}^2   \right\}=   \sum_{n=0}^2 \frac{C_{n}}{2}  \left(\frac{d^{n} \psi}{dt^{n}} \right)^2  = \frac{1}{2} \Psi^\intercal C \Psi,  \label{Lrotpsi}
\end{equation}  
with
\begin{equation}
  C: = S \tilde{C} S^\intercal , \qquad    S= \left(
  \begin{array}{ccc}
  	0 & 0 & 1 \\
  	\frac{2 i C_{2}  }{k \sqrt{2-\frac{8 C_{2} }{k^2}}} & \frac{1}{\sqrt{2-\frac{8 C_{2}   }{k^2}}} &
  	0 \\
  	\frac{2 i C_{1}  }{k \sqrt{2-\frac{8 C_{1}  }{k^2}}} & \frac{1}{\sqrt{2-\frac{8 C_{1}    }{k^2}}} &
  	0 
  \end{array}
  \right) , \qquad
   \Psi:= \begin{pmatrix}
  	\psi \\
  	\psi_t \\
  	\psi_{tt}
  \end{pmatrix} = S \Phi,
\end{equation}  
and $C_n:=C_{nn}$, $S S^\intercal = 1$. The Euler-Lagrange equation resulting from (\ref{Lrotpsi}) is then 
\begin{equation}
	 \sum_{n=0}^2  C_n (-1)^n  \frac{d^{2n} \psi}{dt^{2n}} =0 .
\end{equation}
Using the mode expansion
\begin{equation}
 \psi = \sum_{j=1}^2  a_j e^{-i \omega_j t} + b_j e^{i \omega_j t} ,    \label{modeex}
\end{equation} 
the equation of motion acquires the form 
\begin{equation}
	\sum_{n=0}^2  C_n  (\omega_i)^{2n} =0 . \label{eom59}
\end{equation}
The two canonical fields $\psi_n$ and their corresponding canonical momenta $\pi_n$ computed from (\ref{canrel}) and (\ref{pirel}), respectively, satisfying the canonical Poisson bracket relation  (\ref{canrel})  are then elevated to obey the quantum commutation relations
\begin{equation}
   \left[ \psi_n , \pi_m     \right]  = i \delta_{nm} .
\end{equation}
As verified in detail by Weldon  \cite{weldon03quant}, these relations hold provided the Fock space generators $a_i,b_i$ commute to the residues $R_i$ 
\begin{equation}
	\left[ a_i, b_j  \right]  = \delta_{ij} R_i, \qquad \text{with} \quad  R_i:= \lim_{z \rightarrow \omega_i}    \frac{z-\omega_i}{P(z)}   = \frac{1}{P'(\omega_i)}, \quad P(z) := \sum_{n=0}^2 C_n (z)^{2n} .
\end{equation}
The Hamiltonian $H_k^r$ for the fields $\psi$ following from Ostrogradsky's scheme for scalar field theories as laid out in section 2.2 becomes 
\begin{equation}
	H_k^r= -\frac{1}{2} \left(  C_0 \psi^2 + C_1 \psi_t^2  + 2 C_2 \psi_t \psi_{ttt} - C_2 \psi_{tt}^2       \right) ,
\end{equation}
with $\pi_1 = C_1 \psi_t -C_2 \psi_{ttt}$, $\pi_2 = C_1 \psi_{tt}$, $\Phi_1 = \psi$, $\Phi_2= \psi_t$.   Using the mode expansion (\ref{modeex}) the Hamiltonian can be cast into the form 
\begin{equation}
H_k^r= \frac{1}{2} \sum_{i=1}^2   \frac{\omega_i}{R_i}     \left(   a_i  b_i + b_i a_i    \right) .
\end{equation}
To illustrate the dilemma of HTDT let us first discuss the real scenario for which we assume that $b_i = a_i^\dagger$. Then one may define the vacuum state with energy $E_0 =(\omega_1 + \omega_2 )/2$ by the relation $a_i \left\vert   vac \right\rangle =0 $, following from $ H  \left\vert   vac \right\rangle =E_0 \left\vert   vac \right\rangle $. The norm of the one-particle state $a_i^\dagger   \left\vert  vac \right\rangle $
\begin{equation}
         \left\langle  vac \right\vert   a_i a_i^\dagger   \left\vert  vac \right\rangle  = R_i   ,
\end{equation}
with energy $\omega_i + E_0$ can therefore only be positive when the residue $R_i$ is positive. The energy of an n-particle state constructed from Fock states is $n \omega_i + E_0$. Thus for a theory with positive norm states and a spectrum that is bounded from below we require $R_i >0$ and $\omega_i > 0$. However, as argued in  \cite{weldon03quant} one may only satisfy here one or the other inequality.

Before discussing $L^r_k$, we illustrate this feature with a simpler toy model. Considering for instance a Lagrangian of the form (\ref{Weldonform}) 
\begin{equation}
	   L_t   = \frac{1}{2}   \left[ - \psi^2  + \left( 1+k^2 \right)  \psi_t^2 - \frac{1}{4}   \left( 1+k^2 \right)  \psi_{tt}^2     \right]  ,
\end{equation}
the corresponding equations of motion (\ref{eom59}) has the solutions
\begin{equation}
	  \omega_1^\pm = \pm \sqrt{ 2 +  \frac{2 \vert k \vert}{\sqrt{1+k^2}}}, \qquad  \omega_2^\pm = \pm \sqrt{ 2 -  \frac{2 \vert k \vert}{\sqrt{1+k^2}}}  ,
\end{equation}
where $\omega_i^+ = -\omega_i^- > 0 $. Either two of these solutions may play the role of the $\omega_j$ in (\ref{modeex}). For the related residues we find
\begin{equation}
	R_1^\pm = - \frac{1}{2 \omega_1^\pm \sqrt{k^4+k^2}}, \qquad 
	R_2^\pm =  \frac{1}{2 \omega_2^\pm \sqrt{k^4+k^2} },
\end{equation}
where $R_1^+ =-R_1^- <0$ and $R_2^+ =-R_2^- >0$. Thus we may either pick two states with positive definite norm states and positive residues $R_1^-$, $R_2^+$ at the cost of having an unbounded spectrum from below since $\omega_1^- <0$ or we can require boundedness from below with energy states build from $ \omega_1^+ $, $ \omega_2^+ $ at the prize of having one negative norm state in the theory.

\begin{figure}[h]
	\centering         
	\begin{minipage}[b]{0.53\textwidth}      
		\includegraphics[width=\textwidth]{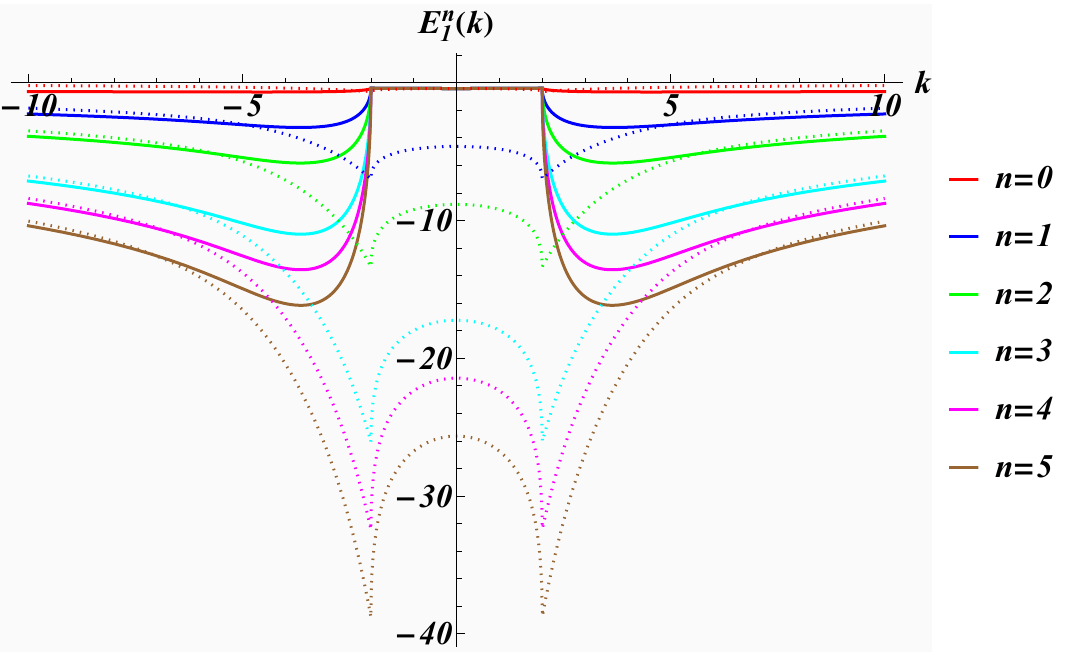}
	\end{minipage}   
		\begin{minipage}[b]{0.46\textwidth}      
		\includegraphics[width=\textwidth]{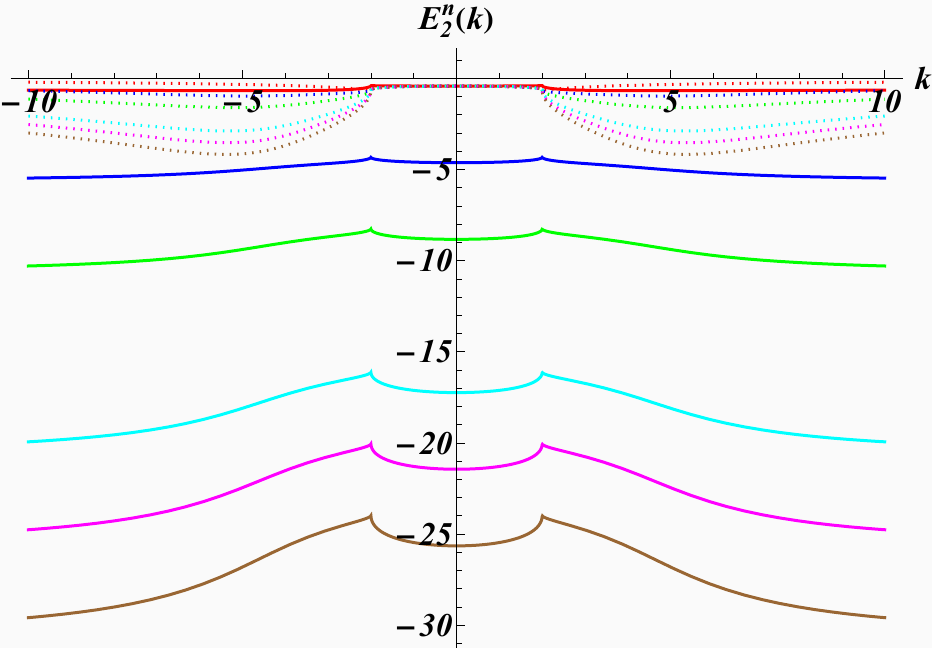}
	\end{minipage} 
	\caption{Quantum spectrum of the $n=2$ rotated modified KdV equation.} 
	\label{spectra}
\end{figure}

Returning to the discussion of the spectrum of $H_k^r$, we find that the corresponding equations of motion (\ref{eom59}) has the solutions
\begin{eqnarray}
	\omega_1^\pm &=& \pm  2 \sqrt{\frac{1}{2 -\sqrt{4-k^2}-\sqrt{24-k^2+4 \sqrt{4-k^2}}}} , \qquad \\ \omega_2^\pm &=& \pm   2 \sqrt{\frac{1}{2-\sqrt{4-k^2}+\sqrt{24-k^2+4 \sqrt{4-k^2}}}} .
\end{eqnarray}
All of these energies are complex with non-vanishing complex part throughout the entire range of $k \in \mathbb{R}$. These energies correspond to resonances with energies $E=E_r - i \Gamma/2$, where $E_r \in \mathbb{R}$ corresponds to the measurable real part of the energy and $\Gamma \in \mathbb{R}^+$ to the decay width. Thus we need to identify the $\omega_j$ in (\ref{modeex}) in such a way that the imaginary parts of the energies are negative for the entire spectrum. The only possible choice that achieves this is to define $E_i^n := E_0-n \omega_i  $ with $E_0:= - \omega_1 - \omega_2$. None of these spectra is bounded from below, as is also illustrated in figure \ref{spectra}. Note that any other choice of the $\omega_i$ would lead to spectra that do not have a positive decay width at least in part of their spectra.

\section{Conclusions}

The Lorentz invariant formulation of Ostrogradsky's method generalised to scalar field theories produced a consistent set of Hamilton's equations when applied to space-time rotated mKdV systems, but it failed to yield the correct time evolution or a conserved Hamiltonian. Instead, employing the nonrelativistic scheme achieved the latter outcomes, which is expected considering the lack of Lorentz invariance in the mKdV system. We outlined an alternative general scheme that circumvents Ostrogradsky's method by concealing time-derivatives beyond the first order in newly defined scalar fields. This approach enables the use of conventional methods to derive canonical variables and associated Hamiltonians. We have demonstrated that mKdV systems can be formulated equivalently in this alternative way. However, it remains an open question whether this equivalence can always be established. The conversion relies on the computation of the first two integrals in (\ref{intconstdes}), and in particular when the $t$-integration can not be carried out trivially, the procedure becomes technically very difficult and may even become impossible to be carried out. In our examples this was the case for the next highest charge not reported here. 

We derived the time-evolution and have constructed consistent sets of equal-time commutation relations for the canonical coordinate and momentum fields of the mKdV systems for generic values of $n$, which should aid future quantisation procedures of the theories. For the special cases of $n=3$ and $n=4$ we also derived the alternative multi-field version of the theories and the consistent time-evolution for the canonical fields of the first higher charge Hamiltonian. These type of Hamiltonians were constructed from inverse Legendre transformed higher charges, which were subsequently space-time rotated before converted into Hamiltonians by means of the Ostrogradsky method. By carrying out a Painlev\'e test we established that the systems obtained in this way are integrable.   

For the $n=2,3,4$-theories we derived several exact real as well as complex benign and malevolent solutions. The classical energies of these systems were shown to be real when the solutions respect a generalised ${\cal CPT}$-symmetry and occur in complex conjugate pairs when that symmetry is broken. By including multi-soliton solutions we showed that one may generate a large set of initial value functions for the Cauchy problem and in the case of $n=4$ we also extended our investigations to the initial-boundary value problem, revealing benign solutions. It remains an open issue whether these type of boundary value problems can be solved in complete generality for the systems considered. Nonetheless, we established that in their functional parameter space of initial boundary value functions they certainly possess islands of benign nature.   However, even though the solutions turned out to be benign, they are unstable as revealed by comparing the exact with the numerical solutions. The nature of the instability differs between the $n=3$, rotated standard KdV-equation, and the $n=4$-case, rotated standard modified KdV equation. In the former model the instability is of a malevolent nature whereas in the latter it is benign, see figures \ref{exnumKdV} and  \ref{exnummKdV} .  Our findings are in agreement with the observations made in \cite{Smilgaacta}. 

We have quantised the $n=2$-theory and for a special toy model were led to the usual dilemma in HTDT that we have to make a choice between two physically undesirable features without being able to avoid both at the same time. We have either a theory with proper normalisable states of positive norm, at the cost of having a spectrum that is not bounded from below, or the reverse, allowing for negative norm states in the system would lead to a spectrum that is bounded from below. In our $n=2$-mKdV theory the choice was dictated by selecting the correct sign in the decay width, leading to a theory that is unbounded from below.  We leave the quantisation of the  $n \neq 2$-theories for future work,  where we also hope to gain new insights into the resolution of the above mentioned fundamental issues of HTDT possibly along the lines mentioned in [8-12].    

\medskip
\noindent \textbf{Acknowledgments:}  TT is supported by JSPS KAKENHI Grant Number 22KJ0752. BT is supported by a City, University of London Research Fellowship.  We would like to thank Andrei Smilga for useful comments.
\appendix 

\section{Superposition principle for the solutions of the rotated KdV}
In this appendix we derive a generalised superposition principle for the solutions of the rotated KdV equation based on B\"acklund transformations and apply it to our solutions (\ref{rotsolKdV}).  Our starting point consists of two different solutions to the rotated time-independent Schr\"odinger equation with different energies $E_{1}$, $E_{2}$ for the same potential $V_0$
\begin{equation}
	-\left(\psi_1 \right)_{tt} + V_0 \psi_1 = E_1 \psi_1 , \qquad -\left(\psi_2 \right)_{tt} + V_0 \psi_2 = E_2 \psi_2 .
	\label{Schrv0}
\end{equation} 
It is then easy to see that the two functions $\hat{\psi}_{1} = 1/ \psi_{1}$, $\hat{\psi}_{2} = 1/ \psi_{2}$ satisfy the space-independent Schr\"odinger equations
\begin{equation}
-\left(\hat{\psi}_1 \right)_{tt} + \hat{V}_1 \hat{\psi}_1 = E_1 \hat{\psi}_1 , \qquad -\left(\hat{\psi}_2 \right)_{tt} + \hat{V}_2 \hat{\psi}_2 = E_2 \hat{\psi}_2 ,
\end{equation} 
with potentials
\begin{equation}
\hat{V}_i = V_0 -2 \left(  \ln\psi_i   \right)_{tt} , \qquad  i=1,2.    \label{pothat}
\end{equation} 
So far the above are simply the initial steps of a rotated version of a Darboux transformation, see e.g. \cite{matveevdarboux}. The connection to the solutions of the KdV equation is made by recalling that the standard Lax operator for the KdV system $L=-\partial_x^2 - u $ can be interpreted as a Hamilton operator with potential $V=-u$ and $u$ being a solution of the KdV equation. This is trivially adapted to the current situation with $x \leftrightarrow t$.  

Introducing next the quantities $u_0 = -V_0$, $u_1 = -\hat{V}_1$, $u_2 = -\hat{V}_2$ and $u_i = (w_i)_t$ the equations above yield
\begin{eqnarray}
	 u_0 + u_1 &=& - \frac{1}{2} \left(w_0 -w_1  \right)^2 - 2 E_1,  \label{A4} \\
	  u_0 + u_2 &=& - \frac{1}{2} \left(w_0 -w_2  \right)^2 - 2 E_2.   \label{A5} 
\end{eqnarray}	
In the spirit of Bianchi’s theorem of permutability \cite{Bianchi}, we can set up a Lamb-diagram \cite{Lamb} by assuming further that there exists a function $u_{12} $ with $u_{12} = (w_{12})_t$ satisfying 
\begin{eqnarray}
	u_1+ u_{12} &=& - \frac{1}{2} \left(w_1 -w_{12}  \right)^2 - 2 E_2,  \\
	u_2 + u_{12} &=& - \frac{1}{2} \left(w_2 -w_{12}  \right)^2 - 2 E_1.
\end{eqnarray}	
We can then combine these four equations to the nonlinear version of the superposition principle
\begin{equation}
	w_{12} = w_0 -4 \frac{E_1 -E_2}{w_1 - w_2}, \label{superpos}
\end{equation} 
so that
\begin{equation}
	u_{12} = u_0 -4 \frac{(E_1 -E_2)(u_1-u_2)}{(w_1 - w_2)^2}. \label{superposu}
\end{equation} 

Next we specialise (\ref{superpos}) to our concrete case. We notice that our solutions (\ref{rotsolKdV}) can be interpreted as rotated version of a shifted Lam\'e potential. Adapting the solutions for such potentials reported in \cite{dunne1998self,CCFsineG}, it is convenient to define the functions $H $, $\Theta $,  $Y$ and $Z$ 
\begin{equation}
	H\left( z\right) :=\vartheta _{1}\left( z\kappa ,q\right) ,\quad \Theta
	\left( z\right) :=\vartheta _{4}\left( z\kappa ,q\right) ,\quad Y(z):= 
	\frac{H_z\left( z \right) }{H\left( z \right) },\quad
	Z(z):= \frac{\Theta_z \left( z \right) }{\Theta \left(
		z \right) },
\end{equation}
defined in terms of Jacobi's theta functions 
$\vartheta _{i}(z,q)$ with $i=1,2,3,4$, $\kappa :=\pi /(2K)$ and nome $q:=\exp
(-\pi K^{\prime} / K)$. Then the combination of the functions
\begin{equation}
      \psi_1^\alpha (t)=  \frac{H(\omega t + p + \alpha) }{\Theta (\omega t + p)}  e^{- (\omega t +p) Z(\alpha)},
      \qquad \text{and} \qquad 
       \psi_2^\alpha (t)=  \frac{\Theta(\omega t + p + \alpha) }{\Theta (\omega t + p)}  e^{- (\omega t +p) Y(\alpha)},  \label{psipsi}
\end{equation}
satisfy the space-independent Schr\"odinger equations (\ref{Schrv0}) for the same potential
\begin{equation}
 V_0 = 2 m \omega^2 \text{sn}^2 (\omega t + p \vert m), 
\end{equation}
with different energies
\begin{equation}
	E_1 = \omega^2 \left[  1+m -m  \text{sn}^2 ( \alpha \vert m)    \right],  \qquad E_2 = \omega^2 \left[  1+m - \text{sn}^{-2} ( \alpha \vert m)    \right],
3\end{equation}
respectively, for $\omega = \omega^{(n)}$ as specified in (\ref{omegan}) and free parameters $\alpha , p$. We have now various options to derive two potentials from the functions in (\ref{psipsi}), either from $ \psi_1^\alpha (t)$ and  $\psi_2^\alpha (t)$, but as the energies depend on $\alpha$ we may also use $ \psi_1^{\alpha_1} (t)$, $ \psi_1^{\alpha_{2}} (t)$ or $ \psi_2^{\alpha_{1}} (t)$, $ \psi_2^{\alpha_{2}} (t)$ with $\alpha_{1}\neq \alpha_{2}$. Here we present the second option. Then specifying $p= k x + \lambda$ we obtain from (\ref{pothat}) the solutions
\begin{equation}
u_i(x,t):=   \left. -2 m \omega^2 \text{sn}^2 (z\vert m) +\left[ \frac{ H^2_{z_i}(z_i)  }{ H^2(z_i)}  
- \frac{ H_{2z_i}(z_i) }{ H(z_i)}  - \frac{ T^2_{z}(z)  }{ T^2(z)}  
+ \frac{ T_{2z}(z) }{ T(z)}  \right] \right\vert_{   { \footnotesize
	\begin{matrix}
	z=k x+ \omega t + \lambda , \\
	z_i =z+ \alpha_i, \qquad  
	\end{matrix}} }
\end{equation}
for $\omega = \omega^{(n)}$. Integrating this over $t$ we obtain
\begin{equation}
	w_i(z)=  \left. -2 \omega \left[ z  -\mathcal{E}(z |m) - \frac{H_{z_i}(z_i)}{H(z_i)  }  + \frac{T_z(z)}{T(z)}   + Z(\alpha_i) \right]  \right\vert_{   { \footnotesize
			\begin{matrix}
				z=k x+ \omega t + \lambda , \\
				z_i =z+ \alpha_i, \qquad  
	\end{matrix}} }
\end{equation}
where $\mathcal{E}(z|m)= \int_0^z  \text{dn}^2 (w  \vert m) dw$ denotes Jacobi's epsilon function. Substituting these expressions into (\ref{superposu}) leads to a superposition with the three free parameters $\lambda$, $\alpha_1 $ and $\alpha_2 $.

\newif\ifabfull\abfulltrue


\begin{thebibliography}{10}
	
	\bibitem{ostrogradsky1850memoire}
	M.~Ostrogradsky,
	\newblock {\em M{\'e}moire sur les {\'e}quations diff{\'e}rentielles relatives
		an probl{\'e}me des isop{\'e}rim{\'e}tres}, volume VI 4,
	\newblock 1850.
	
	\bibitem{Woodard1}
	R.~P. Woodard,
	\newblock Ostrogradsky's theorem on Hamiltonian instability,
	\newblock Scholarpedia {\bf 10}(8), 32243 (2015).
	
	\bibitem{stelle77ren}
	K.~S. Stelle,
	\newblock Renormalization of higher-derivative quantum gravity,
	\newblock Phys. Rev. D {\bf 16}(4), 953 (1977).
		
	
	\bibitem{grav1}
      A.~A. Starobinsky,
    \newblock A new type of isotropic cosmological models without singularity,
    \newblock Phys. Lett. B {\bf 91}, 99-102 (1980).

\bibitem{grav2}
S.~L. Adler,
\newblock Einstein gravity as a symmetry-breaking effect in quantum field theory,
\newblock Rev. Mod. Phys.  {\bf 54}, 729 (1982).

\bibitem{grav3}
A.~Smilga,
\newblock Spontaneous generation of the Newton constant in the renormalizable gravity theory, 
\newblock ITEP preprint 63 (1982) 8 pp, arXiv preprint arXiv:1406.5613 (2014).


	\bibitem{modesto16super}
	L.~Modesto and I.~L. Shapiro,
	\newblock Superrenormalizable quantum gravity with complex ghosts,
	\newblock Phys. Lett. B {\bf 755}, 279--284 (2016).
	
	\bibitem{ghostconst}
	T.-J.~Chen, M. Fasiello, E.~A. Lim, and A.~J. Tolley
	\newblock Higher derivative theories with constraints: Exorcising Ostrogradski's Ghost,
	\newblock JCAP {\bf 2013.02}, 042 (2013).
	
	\bibitem{salvio16quant}
	A.~Salvio and A.~Strumia,
	\newblock Quantum mechanics of 4-derivative theories,
	\newblock The EPJ C {\bf 76}, 1--15 (2016).
	
	\bibitem{fakeons}
	D.~Anselmi,
	\newblock Fakeons and Lee-Wick models,
	\newblock JHEP {\bf 02},  141 (2018).
	
	\bibitem{bender2008no}
	C.~M. Bender and P.~D. Mannheim,
	\newblock No-ghost theorem for the fourth-order derivative Pais-Uhlenbeck
	oscillator model,
	\newblock Phys. Rev. Lett. {\bf 100}(11), 110402 (2008).
	
	\bibitem{raidal2017quantisation}
	M.~Raidal and H.~Veerm{\"a}e,
	\newblock On the quantisation of complex higher derivative theories and
	avoiding the Ostrogradsky ghost,
	\newblock Nucl. Phys. B {\bf 916}, 607--626 (2017).
	
	
	\bibitem{smilga2017rev}
	A.~Smilga,
	\newblock Classical and quantum dynamics of higher-derivative systems,
	\newblock J. Mod. Phys. A {\bf 32}, 1730025 (2017).

	\bibitem{smilga2021exactly}
	A.~Smilga,
	\newblock On exactly solvable ghost-ridden systems,
	\newblock Phys. Lett. A {\bf 389}, 127104 (2021).
	
	\bibitem{Smilga6}
	T.~Damour and A.~Smilga,
	\newblock Dynamical systems with benign ghosts,
	\newblock Phys. Rev. D {\bf 105}, 045018 (2022).
	
	\bibitem{smilga2021benign}
	A.~Smilga,
	\newblock Benign ghosts in higher-derivative systems,
	\newblock in J. of Phys.: Conf. Series, {\bf 2028}  012023 (2021).
	
	\bibitem{Hawking}
	S.~W. Hawking and T.~Hertog,
	\newblock Living with ghosts,
	\newblock Phys. Rev. D {\bf 65}(10), 103515 (2002).
	
	\bibitem{biswas2010towards}
	T.~Biswas, T.~Koivisto, and A.~Mazumdar,
	\newblock Towards a resolution of the cosmological singularity in non-local
	higher derivative theories of gravity,
	\newblock JCAP {\bf 2010}, 008 (2010).
	
	\bibitem{weldon98finite}
	H.~A. Weldon,
	\newblock Finite-temperature retarded and advanced fields,
	\newblock Nucl. Phys. B {\bf 534}(1-2), 467--490 (1998).
	
	\bibitem{mignemi1992black}
	S.~Mignemi and D.~L. Wiltshire,
	\newblock Black holes in higher-derivative gravity theories,
	\newblock Phys. Rev. D {\bf 46}(4), 1475 (1992).
	
	\bibitem{rivelles2003triviality}
	V.~O. Rivelles,
	\newblock Triviality of higher derivative theories,
	\newblock Phys. Lett. B {\bf 577}(3-4), 137--142 (2003).
	
	\bibitem{Kap1}
	D.~S. Kaparulin, S.~L. Lyakhovich, and A.~A. Sharapov,
	\newblock BRST analysis of general mechanical systems,
	\newblock J. of Geo. and Phys. {\bf 74}, 164--184 (2013).
	
	\bibitem{plyush89mass}
	M.~S. Plyushchay,
	\newblock Massless point particle with rigidity,
	\newblock Mod. Phys. Lett. A {\bf 4}(09), 837--847 (1989).
	
	\bibitem{Mpl}
	M.~S. Plyushchay,
	\newblock Massless particle with rigidity as a model for the description of
	bosons and fermions,
	\newblock Phys. Lett. B {\bf 243}(4), 383--388 (1990).
	
	\bibitem{dine1997comments}
	M.~Dine and N.~Seiberg,
	\newblock Comments on higher derivative operators in some SUSY field theories,
	\newblock Phys. Lett. B {\bf 409}(1-4), 239--244 (1997).
	
	\bibitem{smilga17ultrav}
	A.~Smilga,
	\newblock Ultraviolet divergences in non-renormalizable supersymmetric
	theories,
	\newblock Phys. of Part. and Nucl. Lett. {\bf 14}, 245--260 (2017).
	
	\bibitem{Sugg1}
	M.~Pav{\v{s}}i{\v{c}},
	\newblock Stable self-interacting Pais--Uhlenbeck oscillator,
	\newblock Mod. Phys. Lett. A {\bf 28}(36), 1350165 (2013).
	
	\bibitem{Sugg2}
	D.~S. Kaparulin, S.~L. Lyakhovich, and A.~A. Sharapov,
	\newblock Classical and quantum stability of higher-derivative dynamics,
	\newblock EPJ C {\bf 74}, 1--19 (2014).
	
	\bibitem{Sugg3}
	M.~Avendano-Camacho, J.~A. Vallejo, and Y.~Vorobiev,
	\newblock A perturbation theory approach to the stability of the Pais-Uhlenbeck
	oscillator,
	\newblock J. Math. Phys. {\bf 58}(9) (2017).
	
	\bibitem{Sugg4}
	N.~Boulanger, F.~Buisseret, F.~Dierick, and O.~White,
	\newblock Higher-derivative harmonic oscillators: stability of classical
	dynamics and adiabatic invariants,
	\newblock EPJ C {\bf 79}, 1--8 (2019).
	
	\bibitem{deffayet22ghost}
	C.~Deffayet, S.~Mukohyama, and A.~Vikman,
	\newblock Ghosts without runaway instabilities,
	\newblock Phys. Rev. Lett. {\bf 128}(4), 041301 (2022).
	
	\bibitem{deffayet23global}
	C.~Deffayet, A.~Held, S.~Mukohyama, and A.~Vikman,
	\newblock Global and local stability for ghosts coupled to positive energy
	degrees of freedom,
	\newblock JCAP {\bf 2023}(11), 031 (2023).
	
	\bibitem{bethanAF}
	A.~Fring and B.~Turner,
	\newblock Higher derivative Hamiltonians with benign ghosts from affine Toda
	lattices,
	\newblock J. Phys. A: Math. Theor. {\bf 56}, 295203 (2023).
	
	\bibitem{fring23int}
	A.~Fring and B.~Turner,
	\newblock Integrable scattering theory with higher derivative Hamiltonians,
	\newblock Eur. Phys. J. Plus {\bf 138}(12), 1136 (2023).
	
	\bibitem{motohashi20q}
	H.~Motohashi and T.~Suyama,
	\newblock Quantum Ostrogradsky theorem,
	\newblock JHEP {\bf 2020}(9), 1--10 (2020).
	
	\bibitem{Urries}
	F.~J. De~Urries and J.~Julve,
	\newblock Ostrogradski formalism for higher-derivative scalar field theories,
	\newblock J. Phys. A: Math. and Gen. {\bf 31}(33), 6949 (1998).
	
	\bibitem{thibes21nat}
	R.~Thibes,
	\newblock Natural higher-derivatives generalization for the Klein--Gordon
	equation,
	\newblock Mod. Phys. Lett. A {\bf 36}(28), 2150205 (2021).
	
	\bibitem{weldon03quant}
	H.~A. Weldon,
	\newblock Quantization of higher-derivative field theories,
	\newblock Ann. Phys. {\bf 305}(2), 137--150 (2003).
	
	\bibitem{Nutku}
	Y.~Nutku,
	\newblock Hamiltonian formulation of the KdV equation,
	\newblock J. Math. Phys. {\bf 25}, 2007--2008 (1984).
	
	\bibitem{Graham}
	G.~Bowtell and A.~E.~G. Stuart,
	\newblock A particle representation for the Korteweg-de Vries solitons,
	\newblock J. Math. Phys. {\bf 24}, 969--981 (1983).
	
	\bibitem{Miura}
	R.~M. Miura,
	\newblock The Korteweg-de Vries Equation: A Survey of Results,
	\newblock SIAM Review {\bf 18}, 412--459 (1976).
	
	\bibitem{Painor}
	P.~Painlev{\'{e}},
	\newblock M{\'{e}}moire sur les {\'{e}}quations diff{\'{e}}rentielles dont
	l'int{\'{e}}grale g{\'{e}}n{\'{e}}rale est uniforme,
	\newblock Bull. Soc. Math. France {\bf 28}, 201--261 (1900).
	
	\bibitem{ARS}
	M.~Ablowitz, A.~Ramani, and H.~Segur,
	\newblock A connection between nonlinear evolution equations and ordinary
	differential equations of P-type. II,
	\newblock J. Math. Phys. {\bf 21}, 1006--1015 (1980).
	
	\bibitem{Pain1}
	J.~Weiss, M.~Tabor, and G.~Carnevale,
	\newblock The Painlev{\'{e}} property for partial differential equations,
	\newblock J. Math. Phys. {\bf 24}, 522--526 (1983).
	
	\bibitem{Pain2}
	J.~Weiss,
	\newblock The Painlev{\'{e}} property for partial differential equations. II:
	B{\"{a}}cklund transformation, Lax pairs, and the Schwarzian derivative,
	\newblock J. Math. Phys. {\bf 24}, 1405--1413 (1983).
	
	\bibitem{Kruskal}
	M.~Kruskal, N.~Joshi, and R.~Halburd,
	\newblock Analytic and Asymptotic Methods for Nonlinear Singularity Analysis,
	\newblock Lect. Notes Phys. {\bf 638}, 175--208 (2004).
	
	\bibitem{Gramma}
	B.~Grammaticos and A.~Ramani,
	\newblock Integrability- and How to detect it,
	\newblock Lect. Notes Phys. {\bf 638}, 31--94 (2004).
	
	\bibitem{liang03ex}
	C.~Liang, X.~Yong, L.~Zhongfei, and H.~Jiahua,
	\newblock The extended Jacobian Elliptic function expansion method and its
	application to nonlinear wave equations,
	\newblock Fizika A {\bf 12}(4), 161--170 (2003).
	
	\bibitem{fring2020BPS}
	A.~Fring and T.~Taira,
	\newblock Complex BPS solitons with real energies from duality,
	\newblock J. of Phys. A: Math. and Theor. {\bf 53}(45), 455701 (2020).
	
    
	\bibitem{Smilgaacta}
	A.~Smilga,
	\newblock Modified Korteweg-de Vries equation as a system with benign ghosts,
	\newblock Acta Pol. {\bf 61}, 190--196 (2022).
	
	\bibitem{wadati74B}
	M.~Wadati,
	\newblock B{\"a}cklund transformation for solutions of the modified Korteweg-de
	Vries equation,
	\newblock JPSJ {\bf 36}(5), 1498--1498 (1974).
	
	\bibitem{schiff94B}
	J.~Schiff,
	\newblock B{\"a}cklund transformations of MKdV and Painleve equations,
	\newblock Nonlinearity {\bf 7}(1), 305 (1994).
	
	\bibitem{arancibia14sol}
	A.~Arancibia, F.~Correa, V.~Jakubsk{\`y}, J.~M. Guilarte, and M.~S. Plyushchay,
	\newblock Soliton defects in one-gap periodic system and exotic supersymmetry,
	\newblock Phys. Rev. D {\bf 90}(12), 125041 (2014).
	
	\bibitem{nest07inst}
	V.~V. Nesterenko,
	\newblock Instability of classical dynamics in theories with higher
	derivatives,
	\newblock Phys. Rev. D {\bf 75}(8), 087703 (2007).
	
	\bibitem{szegleti20diss}
	A.~Szegleti and F.~M{\'a}rkus,
	\newblock Dissipation in Lagrangian formalism,
	\newblock Entropy {\bf 22}(9), 930 (2020).
	
	\bibitem{sandersdiss}
	J.~Sanders,
	\newblock A dual-oscillator approach to complex-stiffness damping based on
	fourth-order dynamics,
	\newblock Nonlin. Dan. {\bf 109}, 285--301 (2022).
	
	\bibitem{matveevdarboux}
	V.~B. Matveev and M.~A. Salle,
	\newblock Darboux transformation and solitons,
	\newblock (Springer, Berlin)  (1991).
	
	\bibitem{Bianchi}
	L.~Bianchi,
	\newblock Vorlesungen {\"{u}}ber Differentialgeometrie,
	\newblock (Teubner, Leipzig)  (1927).
	
	\bibitem{Lamb}
	G.~L. Lamb~Jr,
	\newblock Analytical descriptions of ultrashort optical pulse propagation in a
	resonant medium,
	\newblock Rev. Mod. Phys. {\bf 43}, 99--124 (1971).
	
	\bibitem{dunne1998self}
	G.~Dunne and J.~Feinberg,
	\newblock Self-isospectral periodic potentials and supersymmetric quantum
	mechanics,
	\newblock Phys. Rev. D {\bf 57}(2), 1271 (1998).
	
	\bibitem{CCFsineG}
	J.~Cen, F.~Correa, and A.~Fring,
	\newblock Degenerate multi-solitons in the sine-Gordon equation,
	\newblock J. Phys. A: Math. Theor. {\bf 50}, 435201 (2017).
	
\end{thebibliography}

\end{document}